\documentclass[aps,prb,twocolumn,superscriptaddress,amsmath,amssymb,footinbib,longbibliography,showkeys]{revtex4-2}
\usepackage{amssymb,amsbsy,amsmath}
\usepackage{graphicx}
\usepackage{dcolumn}
\usepackage{bm}
\usepackage{color}
\usepackage[colorlinks,bookmarks=false,citecolor=blue,linkcolor=red,urlcolor=blue]{hyperref}
\usepackage{xfrac}
\usepackage{wrapfig,tabularx,multirow}
\usepackage{xspace}
\usepackage{mathtools}
\usepackage{microtype}
\usepackage{xcolor}
\usepackage{mathrsfs}
\usepackage{verbatim}
\usepackage[utf8x]{inputenc}	
\usepackage{booktabs}
\usepackage{url}

\newcommand{\KG}{kagom\'e\xspace lattice\xspace}
\newcommand{\KGs}{kagom\'e\xspace lattice \xspace}

\newcommand{\abbref}[1]{Fig.~\protect\ref{#1}}
\newcommand{\figref}[1]{Fig.~\protect\ref{#1}}
\newcommand{\secref}[1]{Sec.~\protect\ref{#1}}
\renewcommand{\eqref}[1]{\text{Eq.}~(\protect\ref{#1})}
\newcommand{\xref}[1]{\protect\ref{#1}}
\newcommand{\fmref}[1]{(\protect\ref{#1})}
\newcommand{\Hi}[1]{\mathcal{H}_{#1}}
\newcommand{\op}[1]{%
    \fontdimen12\textfont3=2pt\fontdimen12\scriptfont3=1.4pt%
    \!\null\mathop{\vphantom{#1}\smash{#1}}\limits_{\sim}\null\!}
\newcommand{\opsc}[2]{\op{s}^{#2}_{#1}}
\def\bra#1{\langle\,{#1}\,|\,}
\def\ket#1{\,|\,{#1}\,\rangle}
\newcommand{\braket}[2]{\langle \, {#1} \, | \, {#2} \, \rangle}

\newcommand{\vek}[1]{{\!\vec{\,#1}}}
\newcommand*{\subt}[1]{_{\text{\scriptsize #1}}}

\newcommand{\vnu}{\vek{\nu}}
\newcommand{\vk}{\vek{k}}
\begin{document}

\title{Non-ergodic one-magnon magnetization dynamics of the kagome lattice antiferromagnet}

\author{Henrik Schl\"uter}
\email{hschlueter@physik.uni-bielefeld.de}
\author{J\"urgen Schnack}
\email{jschnack@uni-bielefeld.de}
\author{Jannis Eckseler}
\email{jeckseler@physik.uni-bielefeld.de}
\affiliation{Fakult\"at f\"ur Physik, Universit\"at Bielefeld, Postfach 100131, D-33501 Bielefeld, Germany}

\date{\today}

\begin{abstract}
    The present view of modern physics on non-equilibrium dynamics is that generic systems 
    equilibrate or thermalize
    under rather general conditions, even closed systems under unitary time evolution. 
    The investigation of exceptions 
    thus not only appears attractive, in view of quantum computing where thermalization is a threat 
    it also seems to be necessary. Here, we present aspects of the one-magnon dynamics on the 
    kagome lattice antiferromagnet as an example of a non-equilibrating dynamics due to flat bands.
    Similar to the one-dimensional delta chain localized eigenstates also called localized magnons 
    lead to disorder-free localization and prevent the system from equilibration.
\end{abstract}
\keywords{Frustrated spin systems, Kagome, Flat bands, Dynamics, Equilibration}
\maketitle

\section{Introduction}
\label{sec-1}

Current theoretical studies on the foundations of thermodynamics focus on the question 
whether a state of equilibration or thermalization is approached in closed quantum systems 
under unitary time evolution.
The path to a deeper understanding has been paved by seminal
works of Deutsch,
Srednicki and many others \cite{Deu:PRA91,Sre:PRE94,ScF:NPA96,Tas:PRL98,RDO:N08,PSS:RMP11,ReK:NJP:12,SKN:PRL14,GoE:RPP16,AKP:AP16,BIS:PR16,WDL:PRB17}. 
In simple terms, generic systems are expected to equilibrate for the vast majority of initial states
and the vast majority of late times. 
In the context of this work, equilibration refers to a long-time behavior where expectation 
values become stationary for the vast majority of late times and assume the same value 
for symmetry-equivalent local operators. The term thermalization is employed if this state corresponds 
to a thermal equilibrium state of some appropriate thermodynamic ensemble
\cite{ScF:NPA96,AKP:AP16,ViR:JSMTE16}.

One way to prevent equilibration is given by the phenomenon of localization
that would slow down or even impede every dynamics.
In Ref.~\cite{JES:PRB23}, the sawtooth chain (also termed delta chain) is examined in the Heisenberg model. 
The system shows for a certain ratio
$J_2/J_1=1/2$ of the two defining exchange interactions
a flat band in one-magnon space or equivalently independent localized one-magnon eigenstates 
of the Hamiltonian -- a phenomenon that has attracted great attention for more than 20 years, 
see e.g.\ \cite{MiT:CMP93,SSR:EPJB01,SHS:PRL02,BlN:EPJB03,RSH:JPCM04,SRM:JPA06,ZhT:PRB04,DRH:PRB10,MHM:PRL12,DRM:IJMP15,LAF:APX18,TDD:ZN20,ROS:PRB22}.
In the context of equilibration, flat bands are interesting because they lead to a vanishing group velocity 
and thus cause a special form of (partial) localization, which is also referred to as \textit{disorder-free localization} \cite{MHS:PRB20}.

In this paper, we examine the \KG, which also has a flat band 
in one-magnon space for homogeneous couplings $J_{ij}=J$. 
We present three types of results: (1) The result of general interest and far reaching importance
is that the mere existence of a flat band prevents the \KG from equilibration. 
For every initial state that contains localized magnons or equivalently parts of the flat band
these contributions will remain time-independent and thus 
inhibit 
equilibration.
We demonstrate that this phenomenon is not an effect of finite size of the model system 
but also holds in the limit 
$N\rightarrow\infty$.
(2) The detailed mathematical argument aims at a decomposition of the investigated state in terms
of one-magnon states localized on hexagons of the lattice which allows an easy interpretation. 
However, the derivation is rather technical in two dimensions
since linear independence and orthogonality is a more subtle issue than in one dimension. 
To obtain rigorous results we introduce a $J_1$-$J_2$-model 
for the \KG with a three-times larger unit cell
which also has a flat band and a reduced number of localized magnons 
related to the flat band. 
The symmetry of the model corresponds to the symmetry of the magnon crystal 
observed close to saturation magnetisation on the \KG \cite{SSH:PRL20}.
This model allows a proper definition both of an approximate orthonormal basis as 
of a strictly orthonormal basis in one-magnon space 
in order to disentangle the contributions to the dynamics of an arbitrary state.
(3) Our findings are visualized by explicit calculations that show the full dynamics 
including stationary and dispersive dynamics.

Although our investigations are performed for a \KG of spins $s=1/2$ the conclusions hold 
for arbitrary spin quantum numbers. In one-magnon space the spin quantum number merely provides
a scaling of the energy.

The paper is organized as follows. In \secref{sec-2} we introduce the system
and some employed concepts. Section~\xref{sec-3} contains the mathematical prerequisites.
Section ~\xref{sec-4} presents the numerical results as well as the interpretation. 
The paper closes with a discussion.

\section{System and Methods}
\label{sec-2}

In the following, we consider the Heisenberg model on the \KG with nearest neighbor antiferromagnetic interaction.
This model possesses $SU(2)$ symmetry, i.e., total spin $S$ and total magnetic quantum number $M$ are good
quantum numbers. The Hilbert space can, e.g., be decomposed into orthogonal subspaces $\Hi{M}$.
We call the subspace with magnetic quantum number $M = S\subt{max}-1$ one-magnon space; its dimension is equal to
the number of spins $N$ of the lattice. It can be described by basis states
        \begin{align}
            \ket{\vec{x}} = 
            \opsc{\vec{x}}{-}\ket{\Omega}\ ,
        \end{align}
where 
        \begin{align}
            \ket{\Omega} = 
            \ket{
                m_{\vec{x}_0}=s,
                \dots,m_{\vec{x}_{N-1}}=s
                }
        \end{align}
denotes the so-called magnon vacuum. The coordinates $\vec{x}$ are given as 
a tuples $\vec{x}=(i, \vnu)$. The number $i$ indicates the position of the spin 
within its unit cell and $\vnu$ indicates the position of the unit cell in 
which the spin is located. States of this basis are in the following 
denoted as $\ket{i, \vnu}$. 
        
Due to translational symmetry this basis of one-magnon space
can be brought into momentum representation
via a Fourier transform, i.e.,
        \begin{align}
            \ket{\chi_i, \vk}=
            \frac{1}{\sqrt{l_1l_2}}
            \sum_{\nu_1=0}^{l_1-1} \sum_{\nu_2=0}^{l_2-1}  
            e^{2\pi\imath\frac{\nu_1 k_1}{l_1}}e^{2\pi\imath\frac{\nu_2 k_2}{l_2}}\underset{= \ket{i, \vnu}}{\underbrace{\op{T}_1^{\nu_1}\op{T}_2^{\nu_2}\ket{i,\vec{0}}}}  \ .
        \end{align}
Since the Hamiltonian is translationally invariant, its eigenvectors can be grouped according to the
momentum quantum number $\vk$ and written as $\ket{\varepsilon_\tau,\vk}$ where $\varepsilon_\tau$ is the 
energy eigenvalue in the energy band $\tau$ for momentum $\vk$.

It turns out, that special energy eigenstates exist for frustrated spin lattices such as the \KG;
these states can be understood in two ways: (1) the states form flat energy bands in momentum representations or
(2) these states represent localized magnons \cite{SSR:EPJB01,SHS:PRL02,DRM:IJMP15}. 
In simple cases such as the 
sawtooth chain they can be transformed into each other via the Fourier transform. 
The localized magnons are labelled as $\ket{\varphi_0, \vnu}$. 
Unstable localized magnons, labelled $\ket{\varphi_\tau, \vnu}$, can also be constructed;
they are related to other bands than the flat band, however, not by a simple Fourier transform.

In this paper, we investigate the time evolution of quantum states.
An initial state
$\ket{\psi_0}=\ket{\psi(t_0)}$ can be evolved by applying the unitary time-evolution operator 
    \begin{align}
        \ket{\psi(t)} = \op{U}(t,t_0=0)\ket{\psi_0}=e^{-\imath \op{H} t/ \hbar}\ket{\psi_0}\ .
    \end{align}
The action of the operator can be evaluated using the eigenstates $\ket{n, \vec{k}}$ 
and the eigenvalues $ E_n(\vec{k})$ of the Hamiltonian that are grouped into bands.
With respect to the time evolved state, the dynamical expectation value of observables can be determined. 
In this paper, we will consider $z$-components of single spin operators $\opsc{i}{z}$ 
re-scaled to values between zero and one:
    \begin{align}
        a_i(t) = s_i -  \bra{\psi(t)}\opsc{i}{z}\ket{\psi(t)}\ , s=1/2 \ .
    \end{align}
    
To usefully examine the propagation of local excitations on the lattice we define
    \begin{align}\label{eq-absenk}
        a(r,t) = \frac{1}{\mathcal{N}(r)}\sum_{r=\mathrm{d}(i, j)} a_i(t), 
        \quad \text{with} \quad \mathcal{N}(r) = \sum_{r=\mathrm{d}(i, j)} 1 \ 
    \end{align}
as a function of the distance $r$ to the initial excitation and the time elapsed 
since this excitation. 
The metric $d(i, j)$ describes the Euklidean distance on the lattice.

\section{Mathematical prerequisites for the \KG}\label{sec-3}

In this section, we explain some mathematical preliminaries necessary to discuss
the dynamics presented in Section~\xref{sec-4}. In particular, we demonstrate how
to decompose a quantum state into a basis that offers a simplified interpretation
of the dynamics since it consists to a large extend of localized magnons. In contrast
to the one-dimensional delta chain, compare Ref.~\cite{JES:PRB23}, issues of linear independence 
make the discussion for the \KG technically more challenging. In particular, 
since the more pictorial basis is non-orthogonal it requires some care when 
interpreting contributions to an arbitrary state since simple projections do not
automatically provide weights.

\subsection{Model for the \KG{}}

\begin{figure}[!ht]
                \centering
                \includegraphics[width=0.7\columnwidth]{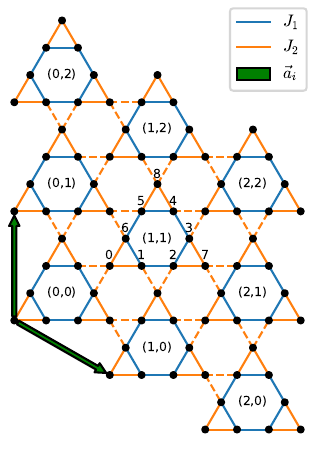}
                \caption{Schematic representation of the \KG{} as a $J_1$-$J_2$ model.
                Dark blue lines denote $J_1$ couplings, light orange both solid 
                and dashed denote $J_2$ couplings.
                The black dots show the positions of the individual spins, 
                the connections between the dots show the corresponding couplings. 
                Solid: coupling between spins of the same unit cell; dashed: 
                coupling between spins of different unit cells, 
                while the couplings of the periodic boundary conditions are omitted for clarity. 
                The unit cells are numbered in the format $(\nu_1,\nu_2)$. 
                The numbering $i=0,\dots,8$ of the spins within a unit cell 
                is shown using the unit cell $(1,1)$ as an example.}
                \label{fig-kago-model-u9}
            \end{figure}

In this paper, the antiferromagnetic \KG is considered within a modified model. 
In contrast to the conventional \KG{} with homogeneous couplings $J_{ij}=J$ between 
nearest neighbor sites $i$ and $j$ and a unit cell of $u=3$ spins and 
correspondingly $l_1 l_2=\tfrac{N}{u}=\tfrac{N}{3}$ unit cells, 
the \KG{} is considered here as a $J_1$-$J_2$ model, with couplings according to 
\figref{fig-kago-model-u9}. This lattice has a unit cell with $u=9$ spins 
and correspondingly $l_1 l_2=\tfrac{N}{9}$ unit cells. 
For the case $J_1=J_2$, the two models agree. 

We fix the units by setting $J_1=1$.
The ratio of the couplings $\alpha = J_2/J_1$ serves as a parameter for the $J_1$-$J_2$ model. 
The differences to the homogeneous model are explained at the relevant points. 
In addition, only grid sections with $l_1=l_2=l$ are considered. 

\subsection{Localized magnons on the \KG{s}}

In one-magnon space, the lattice has exactly as many bands as there are spins in the unit cell.  
The subspace has a dimension equal to the number of spins $N$. As a result, 
the energy eigenstates and eigenvalues are divided into $u=3$ bands for $J_1=J_2$ 
(homogeneous, $\alpha=1$),
while they are divided into $u=9$ bands for $J_1\neq J_2$ (heterogeneous, $\alpha\neq1$). 
The number of eigenstates per band scales accordingly, and so does the Brillouin zone.
                        
Since the antiferromagnetic homogeneous \KG{} with $u=3$ in one-magnon space has a flat band at the lowest energies, 
the \KG{} with $u=9$ in the case $\alpha=1$ has three flat bands with the same energy. 
The $J_1$-$J_2$ model, \figref{fig-kago-model-u9}, is constructed in such a way that its unit cells also fulfill 
the conditions for the existence of localized magnons for different couplings, i.e.\ $\alpha\neq1$, 
cf.~\cite{Schlueter:Diss24}.
                        
\begin{figure}[!ht]
                \centering
                \includegraphics[width=0.99\columnwidth]{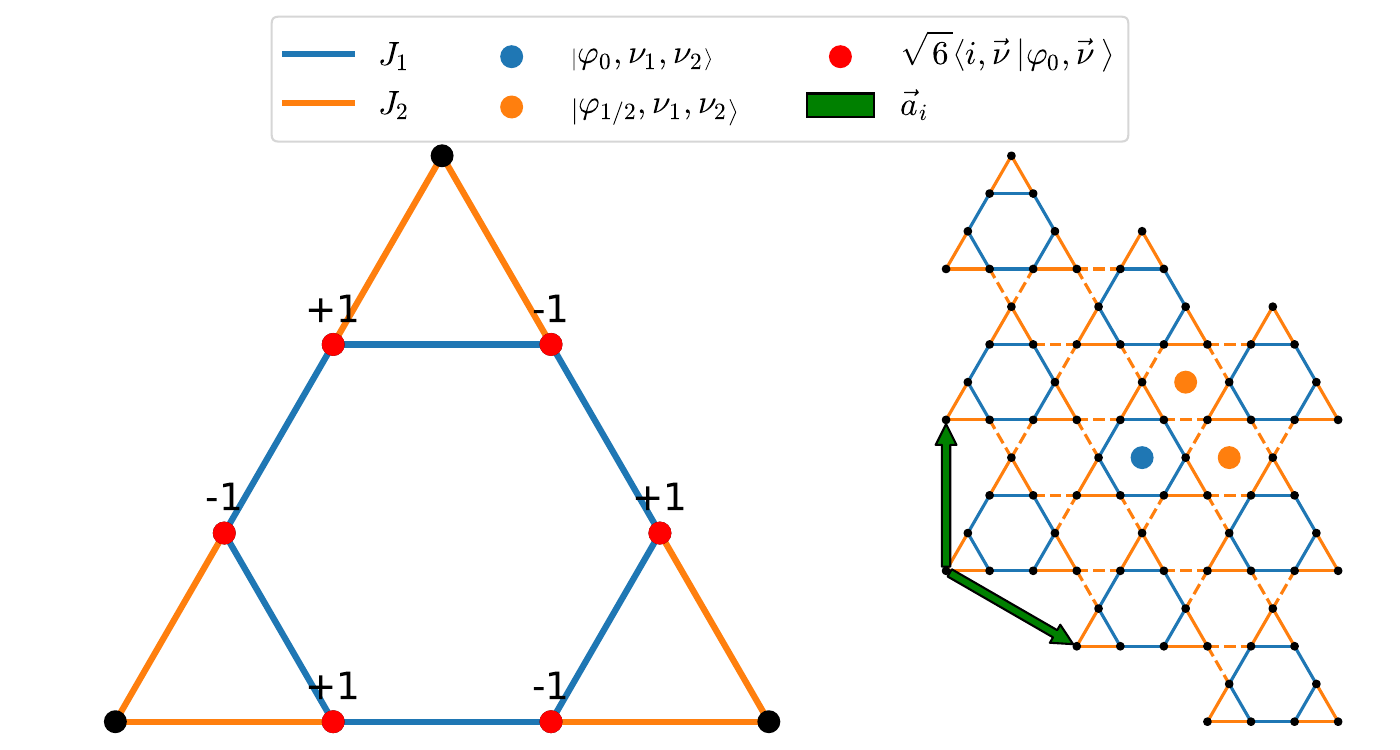}
                \caption{
                    Left: Schematic representation of a localized magnon in a unit cell of the \KG{} 
                    with $u=9$ spin per unit cell. The red dots symbolize the spin sites involved in the state. 
                    The values next to the dots indicate the non-normalized weight and its sign, 
                    i.e.\ the localized magnon is a superposition of one-magnon excitations 
                    from the magnon vacuum at the sites marked in red.
                    Right: For the central unit cell, the hexagon belonging to the stable localized 
                    magnon is marked with a blue dot, while the two hexagons of the unstable localized magnons 
                    are marked with an orange dot.
                }
                \label{fig-kago-log-mag}
            \end{figure}

The number of localized magnons is reduced to one third in the heterogeneous 
case according to the number of unit cells. 
In the limit $\alpha\rightarrow 1$ the $u=9$ model exhibits three localized magnons per unit cell.
The localized magnons, which exist in the case $u=9$ regardless of the value of $\alpha$, can be expressed as 
            \begin{align}\label{eq-loc-mag}
                \ket{\varphi_0 , \vnu} = \sum_{i =1}^6(-1)^i\opsc{i, \vnu}{-}\ket{\Omega}
                \ .
            \end{align}
The sum runs over the indices of the spins of the hexagon in the unit cell, 
see \figref{fig-kago-log-mag} (left).
The hexagon belonging to the stable localized magnons is marked with a blue bullet point in 
\figref{fig-kago-log-mag} (right), 
while the two hexagons giving rise to unstable localized magnons are marked with an orange bullet point. 
            
            \begin{figure}[!ht]
                \centering
                \includegraphics[width=0.85\columnwidth]{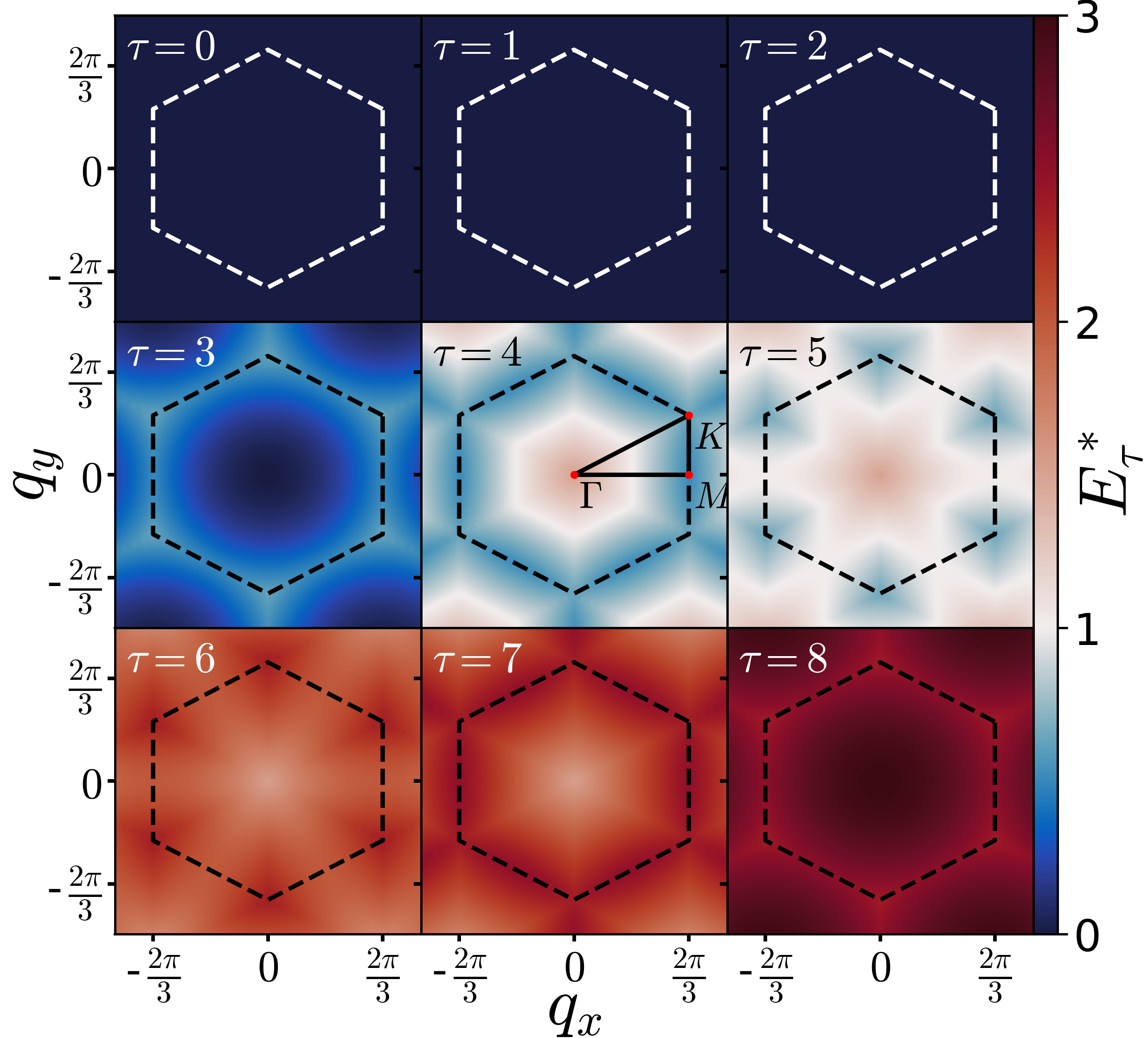}
                \caption{The bands $E^*_\tau$ of the \KG{} with
                  $u=9$ and $\alpha=1$ as a function
                  of the reciprocal vector $\vec{q}\in$1.BZ. The
                  boundaries of the first
                  Brillouin zone can be seen as a dotted
                  hexagon. In the diagram for $\tau=4$,
                  the symmetry points $\Gamma, M$ and $K$ (red
                  dots) and the path between them
                  (black, solid) are also shown, cf. \abbref{fig-kago-band-0-1}.  
                The energy $E^*_\tau$ is the excitation energy above the flat band.}
                \label{fig-kago-band-0}
            \end{figure}
            
\begin{figure}[!ht]
                \centering
                \includegraphics[width=0.99\columnwidth]{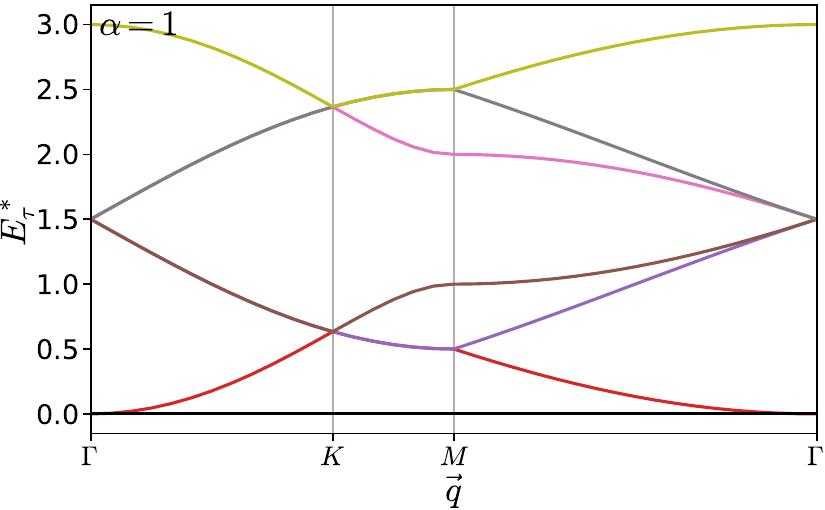}
                 \caption{The bands $E^*_\tau$ of the \KGs{}
                   with $u=9$ and $\alpha=1$ as
                   a function of the reciprocal vectors
                   $\vec{q}\in$1.BZ, which lie on the path
                   between the symmetry points $\Gamma, M$ and $K$, cf. \abbref{fig-kago-band-0}.
                 The energy $E^*_\tau$ is the excitation energy above the flat band.}
                \label{fig-kago-band-0-1}
            \end{figure}

The band structure of the \KGs\ in the $u=9$ model and $\alpha=1$ is shown in Figures \ref{fig-kago-band-0} 
and \ref{fig-kago-band-0-1}. In \figref{fig-kago-band-0-1} the
bands are conventionally shown along the
path between the points of high symmetry. In
\figref{fig-kago-band-0} the bands are shown as a
color plot depending on the components of a vector $\vec{q}\in
\text{rUC}$ (reciprocal unit cell).
The first Brillouin zone is shown as a dashed hexagon in each of
the individual graphs.
In addition, the path (black solid) through the Brillouin zone
with the points of high symmetry (red),
which is used to represent the band structure in
\figref{fig-kago-band-0-1},
is shown as an example in the graph of the central band ($\tau=4$).
             
\begin{figure}[ht!]
                \centering
                \includegraphics[width=0.99\columnwidth]{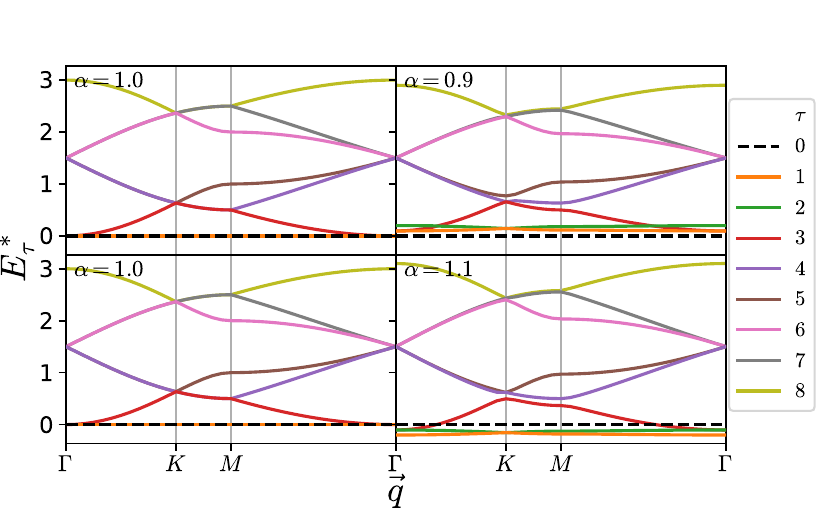}
               \caption{The bands $E^*_\tau$ of the \KGs{} with
                 $u=9$ for different
                 values of $\alpha$ as a function of the
                 reciprocal vectors $\vec{q}\in$1.BZ,
                 which lie on the path between the symmetry
                 points $\Gamma, M$ and $K$,
                 cf. \abbref{fig-kago-band-0}.  
               The energy $E^*_\tau$ is the excitation energy above the flat band.}
                \label{fig-kago-band-1}
            \end{figure}

The three flat bands mentioned above can be seen in the top row of the graphs in \figref{fig-kago-band-0}. 
In \figref{fig-kago-band-0-1}, only one flat line can be seen since the three bands are degenerate. 
Degeneracies between bands begin or terminate at symmetry points.
In the following, we show how the bands deform and degeneracies are lifted 
by comparing the three cases $\alpha=0.9$, $\alpha=1$, and $\alpha=1.1$.
The corresponding band structures are shown in \figref{fig-kago-band-1}. 
The two cases $\alpha=0.9$ and $1.1$ (right) are contrasted with the case $\alpha=1$ 
(shown twice on the left). 
One of the flat bands is shown as a black-dashed line.    

While this band remains flat for all cases, the other two previously flat bands deform 
and move upwards in the case $\alpha=0.9$ 
and downwards in the case $\alpha=1.1$ compared to the flat band. 
In the region between $K$ and $M$ the degeneracy between the red band 
($\tau=3$) and the violet band ($\tau =4$) 
is lifted for both cases with $\alpha\neq 1$.
We define that $\tau=0$ always denotes the remaining flat band, 
even if this no longer contains the ground state. 
The almost flat bands $\tau=1,2$ can be defined by having a 
curvature small compared to the curvature of the remaining bands with $\tau>2$.
The remaining bands are colored in \figref{fig-kago-band-1}
and numbered according to the level of their energy at each value of $\vec{q}$.

\subsection{Possible basis in one-magnon space}
\label{sec-kago-decomp}

The goal of this section is to first construct a non-orthogonal basis that explicitly contains 
all stable
(belonging to the flat band) and non-stable 
(belonging mostly to the almost flat bands)
localized magnons.
Subsequently, an orthonormal basis is worked out, 
which explicitly contains only the stable localized magnons.
Degeneracies between different bands prevent an unambiguous assignment of 
the eigenstates to the bands. For example, for $\alpha=1$ at the $\Gamma$ point, 
it is not possible to determine 
numerically which states belong to the flat bands and which belong to the lowest dispersive band ($\tau=3$). 
The known localized magnon states can be used to partially isolate the states of the three (almost) flat bands.
          
For $\alpha = 1$, it follows from translational symmetry that all localized magnons 
$\ket{\varphi_\tau, \vec{\nu}}$ are completely contained in the three flat bands. 
In the case of $\alpha\neq1$, this applies exclusively to the remaining flat band and 
the stable localized magnons $\ket{\varphi_0, \vec{\nu}}$. 
For the states $\ket{\varphi_{1/2}, \vec{\nu}}$ belonging to the almost 
flat bands this cannot be assumed a priori. 
However, for $\alpha$ close to one it can be assumed that the contribution 
of the states $\ket{\varphi_{1/2}, \vec{\nu}}$ to the almost flat bands is large.

\paragraph{Construction of a basis}
            
When constructing a basis, it must be noted that the localized magnons 
do not represent a complete basis of the flat bands ($\alpha=1$).  
The set of all localized magnons is given by
            \begin{align}
                \mathcal{B}^{\varphi} = \bigcup_{\tau=0}^{2}\mathcal{B}_\tau^\varphi\
            \end{align}
where the quantities $\mathcal{B}^\varphi_\tau $ are given by
            \begin{align}
                \mathcal{B}^\varphi_\tau 
                & = \left\lbrace
                \ket{\varphi_\tau,\vnu}\;\vert\; \vnu
                =(\nu_1, \nu_2): \nu_i = 0,\dots,l-1\right\rbrace
                \\
                & = \left\lbrace \ket{\varphi_\tau,\vnu}\right\rbrace_{\vnu}
                \nonumber
                \ ,
            \end{align}
where the second line is a short hand notation do denote that $\vnu$ assumes 
all possible values in this set.
The set $\mathcal{B}^\varphi$ has a rank \footnote{The rank of a subset 
$A\subseteq V$ of a vector space $V$ means the dimension of the span of $A$.} 
of $\textbf{rank}(\mathcal{B}^\varphi) = 3 l^2 - 1$ \cite{SRM:JPA06}.
Since the localized magnons $\ket{\varphi_\tau,\vnu}\in \mathcal{B}_\tau^\varphi$ 
are pairwise orthogonal for the same $\tau$, cf. \eqref{eq-loc-mag}, 
but the corresponding sets $\mathcal{B}_\tau^\varphi$ have a 
full rank \footnote{A full rank here means that the power of a subset 
$A\subseteq V$ of a vector space $V$ is equal to the dimension of its span, 
i.e., \textbf{rank}$(A) = |A|$ and $A$ is a basis of $\textbf{span}(A)$.}, 
i.e., $\textbf{rank}(\mathcal{B}_\tau^\varphi)=l^2$.
The linear independence is therefore only lost when the three sets are combined.
            
In order to represent states by the localized magnons we exploit the exchange theorem of 
Steinitz \cite{GF:SFW13} according to which a basis with the same linear span 
can be found by iteratively removing individual states from a generating system.
Thus, if the set of all localized magnons $\mathcal{B}^\varphi$
is reduced by a single localized magnon $\ket{\varphi_\tau, \vnu_1}$ to form the set
            \begin{align}
                {\mathcal{B}^\varphi}^\prime = \mathcal{B}^\varphi\ 
                \backslash\ \lbrace\ket{\varphi_\tau, \vnu_1}\rbrace\ ,
            \end{align}
which has full rank. Therefore, ${\mathcal{B}^\varphi}^\prime$ is a basis of the span of the localized magnons. 
            
The set of all energy eigenstates $\ket{\varepsilon_\tau,\vk}$ with $\tau > 2$ is given by
            \begin{align}
                \mathcal{B}^\varepsilon = \bigcup_{\tau=3}^{u-1}\mathcal{B}_\tau^\varepsilon\ ,
            \end{align}
            where $\mathcal{B}_\tau^\varepsilon$ are defined as
            \begin{align}
               \mathcal{B}_\tau^\varepsilon = \left\lbrace \ket{\varepsilon_\tau,\vk}\right\rbrace_\vk\ .
            \end{align}
Since the basis ${\mathcal{B}^\varphi}^\prime$ by construction spans part of the subspace of 
the three (almost) flat bands and is thus linearly independent of 
$\mathcal{B}^\varepsilon$ \footnote{Here it is crucial that the states are sorted 
according to the above criteria of their eigenvalues.}, the following applies 
            \begin{align}
                \textbf{rank}(\mathcal{B}^{\varphi\prime}\cup
                \mathcal{B}^\varepsilon)
                = \textbf{rank}(\mathcal{B}^{\varphi\prime}) + \textbf{rank}(\mathcal{B}^\varepsilon) = D-1 \ .
            \end{align}
However, due to the degeneracies between the bands, states $\ket{\psi_\varepsilon}\in\mathcal{B}^\varepsilon$ 
are not \textit{a priori} orthogonal to states $\ket{\psi_\varphi}\in\mathcal{B}^{\varphi\prime}$. 
With the aim of constructing a basis that enables the representation with respect to the localized magnons, 
the parts of the basis ${\mathcal{B}^\varphi}^\prime$ can be determined from the states 
$\ket{\psi_\varepsilon}\in\mathcal{B}^\varepsilon$ using
            \begin{align}
                \label{eq-14}
                \ket{\psi_\varepsilon^{\prime\prime}} &=
                \ket{\psi_\varepsilon} -
                \sum_{\ket{\psi_\varphi^\perp}\in\mathcal{B}^{\varphi\prime}_\perp}
                \ket{\psi_\varphi^\perp}\braket{\psi_\varphi^\perp}{\psi_\varepsilon}\ ,\\
                \ket{\psi_\varepsilon^{\prime}} &= \frac{\ket{\psi_\varepsilon^{\prime\prime}}}
                {\braket{\psi_\varepsilon^{\prime\prime}}{\psi_\varepsilon^{\prime\prime}}}
                \label{eq-15}
            \end{align}
where $\mathcal{B}^{\varphi\prime}_\perp$ is an orthogonal version of the base 
$\mathcal{B}^{\varphi\prime}$. 
All states $\ket{\psi_\varepsilon^\prime}$ constructed via Eqs.~\fmref{eq-14} and \fmref{eq-15}
form a new basis $\mathcal{B}^{\varepsilon\prime}$ which, however, must still be re-orthonormalized. 
These states are no longer eigenstates of the Hamilton operator.
            
If one now finds an auxiliary state $\ket{\xi}\in
\Hi{S\subt{max}-1}$ which is simultaneously orthogonal to
${\mathcal{B}^\varphi}^\prime$ and
${\mathcal{B}^\varepsilon}^\prime$,
the bases ${\mathcal{B}^\varphi}^\prime,
{\mathcal{B}^\varepsilon}^\prime$
can be completed with this state to form a basis of the one-magnon space
            \begin{align}
                \mathcal{B} = {\mathcal{B}^\varphi}^\prime \cup \mathcal{B}^{\varepsilon\prime} \cup \{\ket{\xi}\} 
                \ .
            \end{align}
It should be noted that the decomposition of the initial state in this basis strongly depends on the choice of the removed 
localized magnon $\ket{\varphi_\tau, \vnu_1}$. Due to the local
character of the magnons and the excitation,
it can be assumed that it makes sense to choose $\vnu_1$ as far
away as possible from $\vnu_0$.
The observations show that the type $\tau$ of the distant magnon
has no great influence on the decomposition
as long as its position $\vnu_1$ is sufficiently far away from
the excitation.
The contribution of $\ket{\xi}$ is always small 
$\lesssim 10^{-4}$ compared to the total state. 
The type of distant magnon is therefore defined as $\tau=2$.
            
The system of linear equations to be solved can be set up for the constructed basis
which yields 
            \begin{align}
                \braket{i, \vnu}{i_0, \vnu_0} 
                & =
                \sum_{\tau=0}^{1}
                \sum_{\vnu^\prime} 
                \braket{i, \vnu}{\varphi_\tau, \vnu^\prime}\
                x_{\tau, \vnu^\prime} \\
                &+
                \sum_{\vnu^\prime\neq\vnu_1}
                \braket{i, \vnu}{\varphi_2, \vnu^\prime}\ 
                x_{2, \vnu^\prime}+ 
                \braket{i, \vnu}{\xi}\ 
                \gamma_{\xi} \\
                &+
                \sum_{n=0}^{D-3 l^3}
                \braket{i, \vnu}{\psi_{\varepsilon, n}^\prime}\ 
                \beta_{n}
                \ .
            \end{align}
The states $\ket{\psi_\varepsilon^\prime}$ and $\ket{\xi}$ are orthogonal to all base states. 
Accordingly, their components can be expressed with the help of scalar products
            \begin{align}
                \beta_{n} =
                \braket
                {\psi_{\varepsilon, n}^\prime}
                {i_0, \vnu_0} 
                \quad\text{and}\quad 
                \gamma_{\xi} = 
                \braket
                {\xi}
                {i_0, \vnu_0}
                \ .
            \end{align}{}
Finally, the state $\ket{i_0,\vnu}$ can be represented as
            \begin{align}\label{eq-kago-decomp}
                \ket{i_0, \vnu_0} 
                & =
                \sum_{\tau=0}^{1}
                \sum_{\vnu} 
                x_{\tau, \vnu}
                \ket{\varphi_\tau, \vnu}
                +
                \sum_{\vnu'\neq\vnu_1} 
                x_{2, \vnu'}
                \ket{\varphi_2, \vnu'}
                \\
                & + 
                \gamma_{\xi}
                \ket{\xi}
                +
                \sum_{n=0}^{D-3 l^3}
                \beta_{n}
                \ket{\psi_{\varepsilon, n}^\prime}
                \nonumber
                \ .
            \end{align}

\paragraph{Decomposition of a state}
            
A problem of non-orthogonal but normalized bases is that the representation of a state in 
such a basis is generally not normalized, even if the state is normalized in an orthonormal basis. 
This means that it is not possible to talk about the contributions of different magnon types $\tau$ 
without orthogonalizing the basis. However, orthogonalization causes the local character
of the states to be lost. 
It would also no longer be possible to distinguish between the three (almost) flat bands. 
Nevertheless, it is possible to keep a maximum subset of orthogonal states and adapt the remaining states.
            
Since the stable localized magnons $\ket{\varphi_0,\vnu}$ are pairwise orthogonal, 
but not orthogonal to the unstable localized magnons, these form such a maximum subset. 
The contributions of the stable localized magnons can be derived 
after the stable localized magnons $\ket{\varphi_{0},\vnu}$ 
have been projected out
            \begin{align}
                \ket{\varphi_{1/2}^{\prime\prime},\vnu} &= \ket{\varphi_{1/2},\vnu} - \sum_{\vnu} 
                \ket{\varphi_0,\vnu}\braket{\varphi_0,\vnu}{\varphi_{1/2},\vnu}\ ,\\
                \ket{\varphi_{1/2}^{\prime}, \vnu} 
                &= 
                \frac{
                    \ket{\varphi_{1/2}^{\prime\prime},\vnu}
                }{
                    \braket{
                        \varphi_{1/2}^{\prime\prime},\vnu
                    }{
                        \varphi_{1/2}^{\prime\prime}, \vnu
                    }
                } \ .
            \end{align}
The states $\ket{\varphi_{1/2}^{\prime}, \vnu}$ can then be orthogonalized with respect to each other. 
Together with the stable localized magnons $\ket{\varphi_0,\vnu}$, 
the auxiliary state $\ket{\xi}$ and the partial basis $\mathcal{B}^{\varepsilon\prime}$ 
an orthonormal basis of the one-magnon space is finally obtained
            \begin{align}\label{eq-ONB}
                \mathcal{B}^{\varphi_0} = \left\{\ket{\varphi_0, \vnu}\right\}_\vnu 
                \cup \left\{\ket{\varphi_{1/2}^{\prime}, \vnu}\right\}_\vnu \cup \{\xi\}\cup 
                \mathcal{B}^{\varepsilon\prime}
                \ ,
            \end{align}
which contains the stable localized magnons. 
Due to the obtained orthogonality of the basis and the assignment of its states 
to the stable localized magnons, the unstable localized magnons and the remaining basis, 
the parts of the always flat band $P_0$ can be distinguished from the parts of the 
almost flat bands $P_{1,2}$ and these two from the parts of the remaining bands 
$P_{\tau>2}$. 
The contributions can be defined as follows
            \begin{itemize}
                \item[1.] the contribution of the always flat band $\tau=0$
                    \begin{align}
                        P_{0} = \sum_{\vnu} \vert\braket{\varphi_0,\vnu}{i_0,\vnu_0}\vert^2\ ,
                    \end{align}
                \item[2.] the contribution of almost flat bands, i.e., $\tau=1$ and $\tau=2$
                    \begin{align}\label{1/2}
                        P_{1,2} 
                        & = 
                        \sum_{\vnu} \vert\braket{\varphi_1^\prime,\vnu}{i_0,\vnu_0}\vert^2 + 
                        \sum_{\vnu} \vert\braket{\varphi_2^\prime,\vnu}{i_0,\vnu_0}\vert^2 
                        \\
                        & + 
                        \vert\braket{\xi}{i_0,\vnu_0}\vert^2
                        \nonumber
                        \ ,
                    \end{align}
                \item[3.] the contribution of the remaining bands $\tau>2$
                    \begin{align}
                        P_{\tau > 2} = \sum_{n=0}^{D-l^3} \vert\braket{\psi_{\varepsilon, n}}{i_0,\vnu_0}\vert^2\ .
                    \end{align}
            \end{itemize}

\section{Dynamics of the \KG{s}}\label{sec-4}

In this section, we present our numerical results for the dynamics in one-magnon space.
As an observable we investigate the local magnetization at the spin sites which provides
a very convenient means to visualize localization as well as delocalization in the course
of the time evolution.

\subsection{Numerical results in one-magnon space}\label{sec-kago-dynamic}

This section examines the dynamic behavior of a rather large \KG with $N=11,664$ in the $u$=$9$ model. 
This system size corresponds to a linear dimension of $l_1=l_2=l=36$. 
It is indispensable to work with a large system in order to achieve a 
sufficient spatial resolution of the propagation of investigated excitations. 
In addition, the number of $k$ points, which determines the resolution of the bands, 
also increases with system size. 
Since the energies of the bands are not known analytically in this case, 
numerical approximations must be used to determine the gradients needed for the group velocities, 
for example. 
The quality of such an approximation is strongly dependent on the resolution of the available data. 

In our investigation we address two possible initial states, 
namely the local excitations
            \begin{align}
                \ket{i_0, \vnu_0} = \opsc{i_0,\vnu_0}{-}\ket{\Omega}
            \end{align}
with $i_0=0$ and $i_0=1$ which due to symmetry are the only relevant cases for $\alpha\ne 1$.
In addition to the translational symmetry a discrete rotational symmetry 
about $\theta=\tfrac{\pi}{3}$ around the centers of the hexagons is used here
(point group $C_3$).
A site with $i_0=0$ corresponds to a spin that sits at the apex of the triangle of the unit cell and is not directly connected to the stable localized magnon of its unit cell, 
cf. \figref{fig-kago-log-mag} (left). 
A site with $i_0=1$ corresponds to a spin on the hexagon; it belongs to
the stable localized magnon of its unit cell.
            
            \begin{figure}[!ht]
                \centering
                \includegraphics[width=0.99\columnwidth]{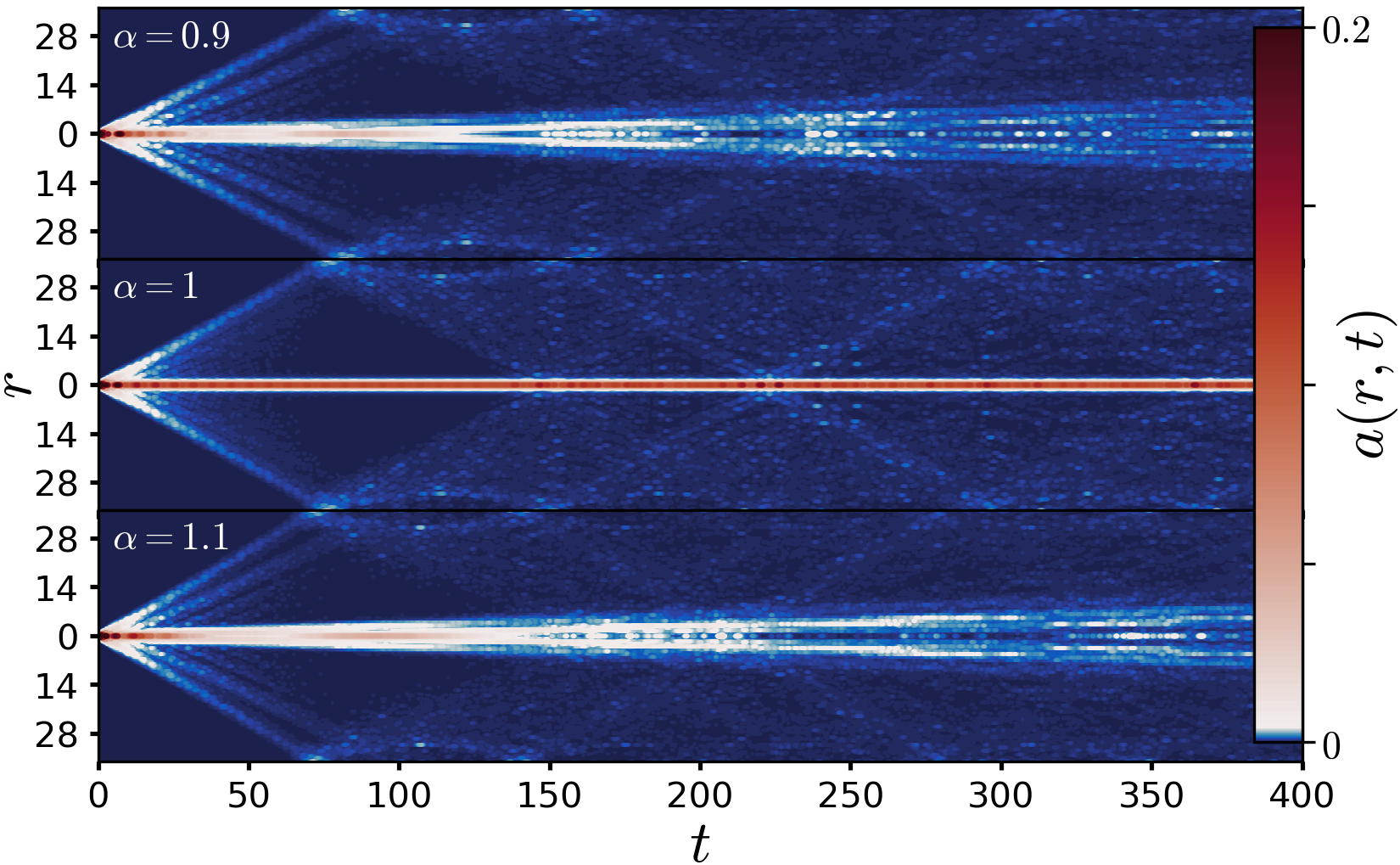}
                \caption{
                Local magnetization $a(r, t)$,
                \eqref{eq-absenk},
                of the \KG with $l=36$ as a function of time and
                distance $r$ to the position of a
                local excitation $\ket{0,\nu_0}$ for different values of $\alpha$. 
                }
                \label{fig-kago-ste}
            \end{figure}

In \figref{fig-kago-ste} the results for the local magnetization $a(r, t)$ are shown 
for the case $i_0=0$ and for different values of $\alpha$. 
The color corresponds to the value of $a(r, t)$ according to the color bar on the
right hand side of the figure. The darkest color corresponds to the polarized background,
lighter colors trace the dynamics of the excitation. Straight features signal 
a dynamics with constant velocity. Due to periodic boundary conditions moving features 
reenter from the opposite side after having left the supercell. 

For $\alpha=1$ one sees a major feature that does not move away from zero;
this is the part of the initial state that belongs to the localized magnon.
One also sees features that propagate very fast away from the point of excitation.
In contrast to the sawtooth ring, this structure disappears after the first orbit,
i.e., after reaching the boundary and reentering. 
There are two possible reasons for this difference to the one-dimensional case.
            \begin{itemize}
                \item[1.] Compared to the results for the
                  sawtooth ring in
                  \cite{JES:PRB23,Schlueter:Diss24},
                  the \KG has more bands, which form a much more
                  complex structure,
                  i.e., a larger number of different speeds are
                  involved.
                  However, the velocities are sufficiently close to each other so that this is not immediately visible. 
                \item[2.] The two-dimensional character of the
                  \KG allows
                  the excitation to reach other places on several different paths, which favors interference. 
                The periodic boundary conditions increase this effect. 
                This results in the complete loss of the structure after the first cycle.
            \end{itemize}
It is therefore plausible that the \KG shows a more complex dynamical behavior than the sawtooth ring.
            
In the other two situations $\alpha\neq1$ it can be seen that further propagation 
velocities play a role right from the start. This can be explained in accordance with point 1. 
As can be seen in \figref{fig-kago-band-1}, the bands for $\alpha\neq1$ 
are degenerate in significantly fewer places. This further increases the number of possible velocities. 
The part of the excitation that remains local for $\alpha=1$ appears to decay completely for the cases 
$\alpha\neq 1$. This can be partly explained by the curvature of the previously flat bands. 
            
To investigate these speculations, for each band $\tau$ the maximum group velocity
            \begin{align}
            \label{maxgrad}
                v\subt{max}^\tau 
                = \max_{\vec{q}}\vert\vert\nabla_{\vec{q}} E_\tau(\vec{q}) \vert\vert
            \end{align}
is determined. 

            \begin{figure}[!ht]
                \centering
                \includegraphics[width=0.99\columnwidth]{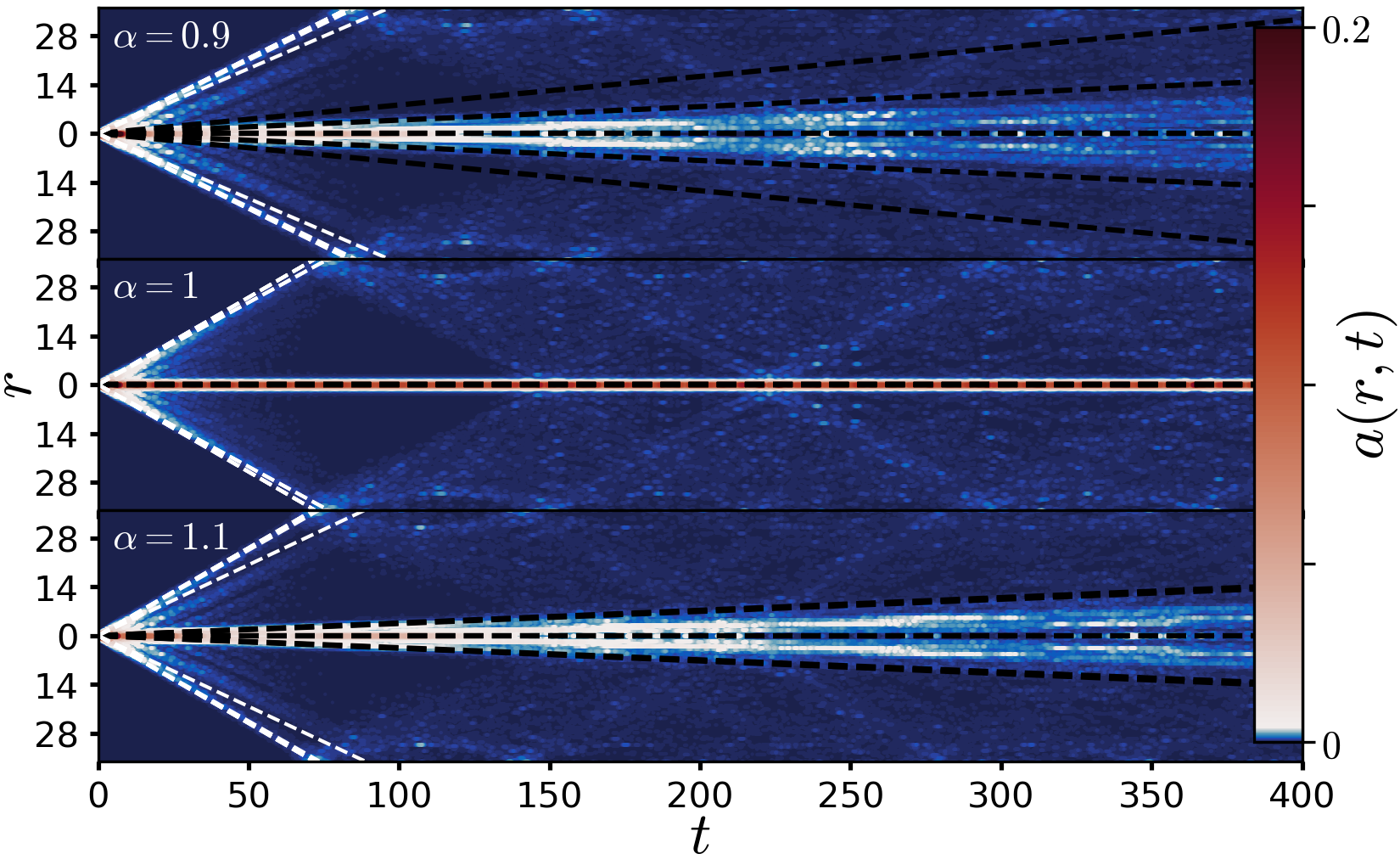}
                \caption{
                The local magnetization $a(r, t)$, \eqref{eq-absenk}, of the \KG with $l=36$ as a function of 
                time and distance $r$ to the position of a local excitation $\ket{0,\nu_0}$ for different values of $\alpha$. 
                In addition, straight dashed lines with slopes corresponding to the maximum gradients $v\subt{max}^\tau$ 
                of the bands are superimposed. These are marked for bands $\tau\leq2$ with black-dashed lines 
                and for bands $\tau>2$ with white-dashed lines.
                }
                \label{fig-kago-ste-1}
            \end{figure}
            
In \figref{fig-kago-ste-1} the straight dashed lines with the corresponding gradients 
according to \fmref{maxgrad}
are superimposed. The straight lines for $\tau > 2$ are shown in white-dashed, 
while the straight lines for $\tau \leq 2$ are shown in black-dashed. 
For all three values of $\alpha$, the straight line of a flat band can be seen;
for $\alpha=1$ the straight lines of the two other flat bands correspond to the first one. 
In the cases $\alpha \neq1$, the straight lines of the almost flat bands match the decay of 
the otherwise remanent component very well, although in the case $\alpha=0.9$ 
no contribution for the slope of the black line with the largest slope can be recognized in the dynamics.
The lines corresponding to the remaining bands confirm the assumption that 
the scattering of the velocities for $\alpha\neq1$ increases compared to the homogeneous grid. 
They seem to fit well with the additional propagation velocities mentioned above, 
although it should be emphasized that only the respective maximum group velocities 
of the bands are shown.
            
\begin{figure}[!ht]
                \centering
                \includegraphics[width=0.99\columnwidth]{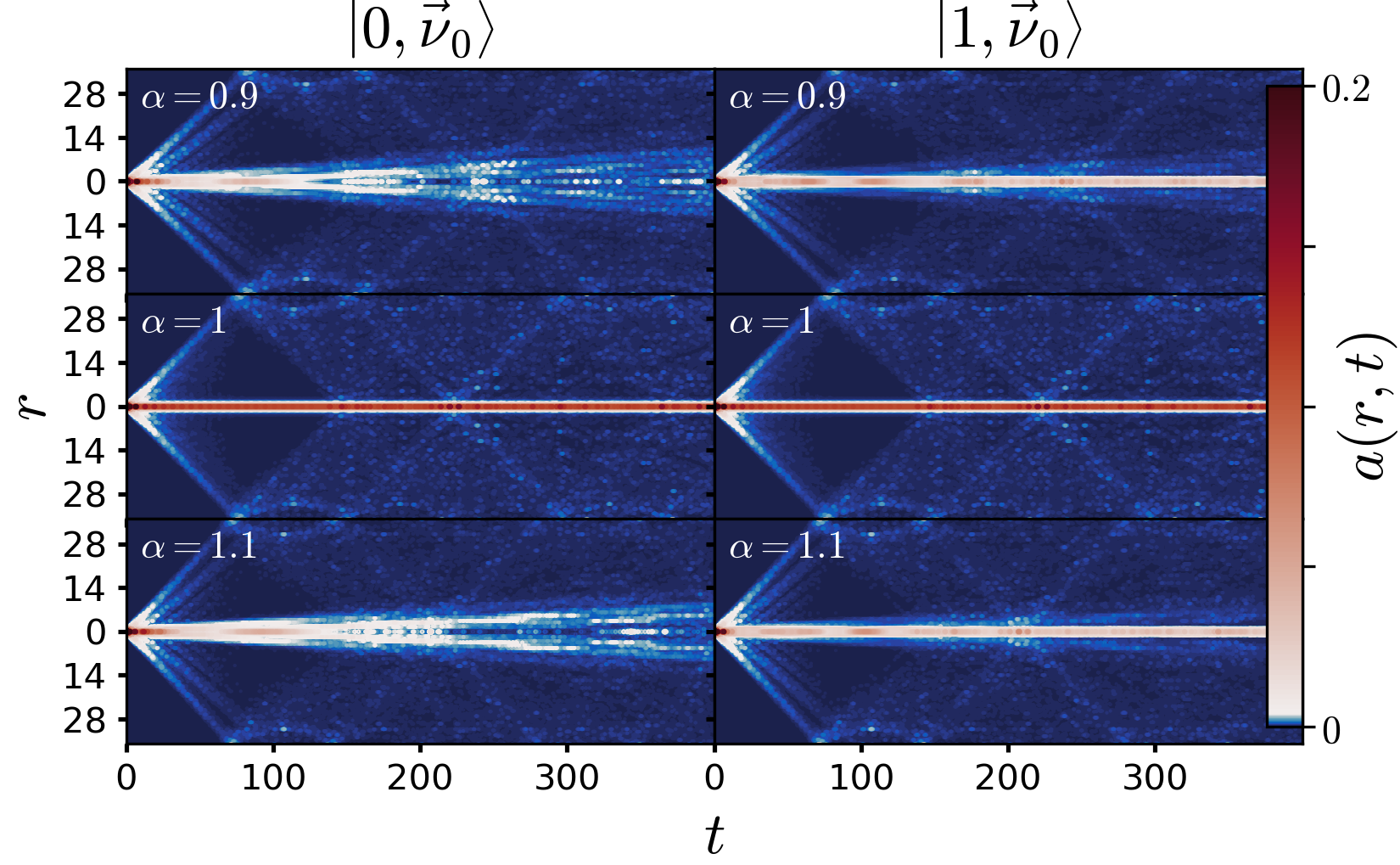}
                \caption{
                The local magnetization $a(r, t)$, \eqref{eq-absenk}, 
                of the \KG with $l=36$ as a function of the time and the distance $r$ 
                to the position of the respective local excitation $\ket{i_0,\nu_0}$ 
                for different values of $\alpha$. 
                The results for $i_0=0$ are shown on the left and for $i_0=1$ on the right.}
                \label{fig-kago-ste-flip}
            \end{figure}

The comparison with the behavior of a spin flip at a place with $i_0=1$ is shown 
in \figref{fig-kago-ste-flip}. It is easy to see that, in contrast to the previous initial state, 
a clear local remanence can be seen for all three values of $\alpha$.
Here, too, there is a slow component in the cases $\alpha\neq1$, 
which is associated with the almost flat bands. 
However, this part of the initial state appears to be significantly smaller than before.  

To clarify why the excitation appears to decay completely in the
situations with $\alpha\neq1$
and for an initial state with $i_0=0$ compared to an initial state with $i_0=1$, the contributions $P_\tau$
of the initial state in the bands are listed in Table \ref{tab:my_label_2}. 
It is noticeable that the contributions of the bands are independent of the value of $\alpha$. 
This is due to the fact that
the subspaces of the bands with $\tau\leq2$ can also be described for 
$\alpha\neq 1$ in terms of local magnons as explained in the previous section.
Since their definition does not change with $\alpha$, 
the representation also remains independent. 

            \begin{table}[!ht]
                \centering
                \begin{tabular}{|c|c|c|c|c|}
                    \toprule
                    $i_0$ & $P_\tau$ & $\alpha = 0.9$ & $\alpha = 1$ & $\alpha = 1.1$\\\hline
                    \multirow{3}{*}{0} & $P_0$ & $0$ & $0$ & $0$\\
                                        & $P_{1,2}$ & $0.333$ & $0.333$ & $0.333$\\
                                        & $P_{>2}$ & $0.667$ & $0.667$ & $0.667$\\\hline
                    \multirow{3}{*}{1} & $P_0$ & $0.167$ & $0.167$ & $0.167$\\
                                        & $P_{1,2}$ & $0.167$ & $0.167$ & $0.167$\\
                                        & $P_{>2}$ & $0.667$ & $0.667$ & $0.677$\\\bottomrule
                \end{tabular}
                \caption{Contributions $P_\tau$ of the bands to the initial states $\ket{i_0,\vnu_0}$.}
                \label{tab:my_label_2}
            \end{table}

For the initial state with $i_0=0$, we find that its share of the always flat band is zero. 
This is consistent with the observation that in a system with $\alpha\neq1$ this state 
is distributed on the lattice over time. For an initial state with $i_0=1$, on the other hand, 
the contribution from the always flat band is one sixth of the entire state, 
which is also consistent with the dynamics. It is important to note that for $\alpha=1$ the 
bands $\tau=1,2$ are also flat, and so the contribution of flat bands in both initial states 
increases to a third. This explains why the localized fraction in the dynamics is 
largest for $\alpha=1$ and also exists for an initial state with $i_0=0$. 
Above all, this is consistent with the expectation that the symmetry between the 
two excitations is restored at $\alpha=1$. Furthermore, the decomposition within 
the systems under consideration is independent of the system size, which confirms 
the expectation that the dynamics is not an effect of finite system sizes either.

\subsection{Interpretation of the dynamics of the \KG{s} by localized magnons}
            
Following the work in \cite{JES:PRB23,Schlueter:Diss24} on the sawtooth ring, in this section
an analysis of the dynamics of the system is carried out using the localized magnons. 
For this purpose, the representation of a state $\ket{i_0,\vnu_0}$ in the non-orthogonal 
basis $\mathcal{B}$ developed in section \ref{sec-kago-decomp} is used. 
The essential components of the decomposition $\vert x_{\tau, \vnu}\vert^2$ 
according to \fmref{eq-kago-decomp} are considered.
The components of the dispersive bands $\vert \beta_{n}\vert^2$, 
on the other hand, are already analyzed in section \ref{sec-kago-dynamic}. 
            
            \begin{figure}[!ht]
                \centering
                \includegraphics[width=0.99\columnwidth]{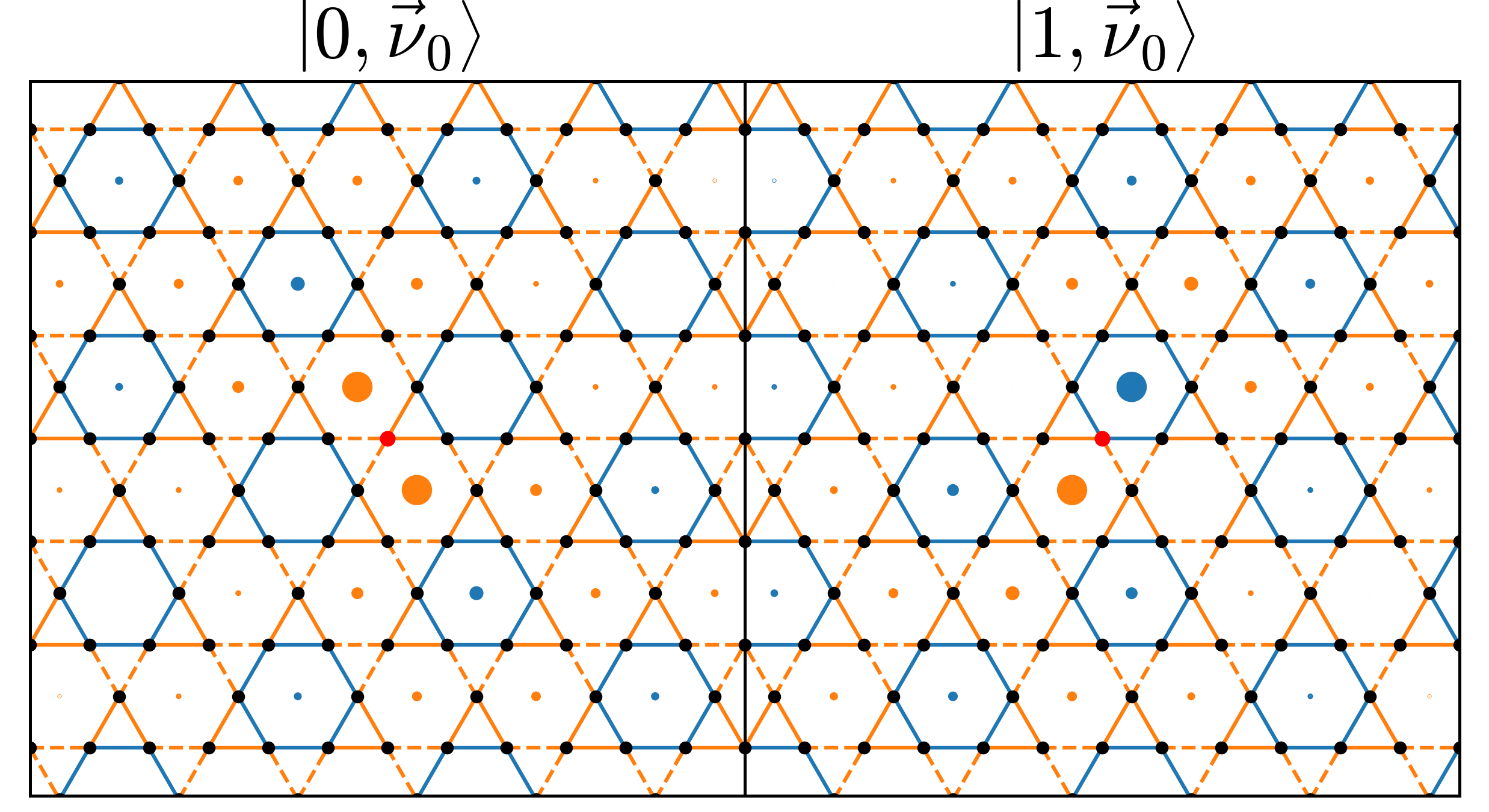}
                \caption{
                    Schematic representation of the decomposition $\vert x_{\tau, \vnu}\vert^2$ 
                    of the two initial states into localized magnons, cf. \eqref{eq-kago-decomp}. 
                    The hexagons belonging to the stable localized magnon are marked with blue dots, 
                    while the hexagons of the unstable localized magnons are marked with orange dots. 
                    The size of the dots indicates the magnitude squared of the amplitude of the respective 
                    localized magnon. The position of the excitation $\ket{i_0,\vnu_0}$ is marked 
                    with a red dot in the center of each structure.
                }
                \label{fig-kago-decomp}
            \end{figure}
            
Figure \xref{fig-kago-decomp} shows the contributions of the localized magnons to the initial states. 
The size of the points corresponds to $\vert x_{\tau, \vnu}\vert^2$. 
The position of the excitation is marked with a small red dot in the center of each picture. 
As already explained, for $i_0=0$ the contributions of the stable localized magnons 
to the initial state are small (small points in \figref{fig-kago-decomp} (left)), 
while the two largest components belong to unstable localized magnons
(bigger orange points in \figref{fig-kago-decomp} (left)).
The small but finite components of the stable localized magnons in this state 
do not contradict the value of $P_0=0$ in Tab.~\ref{tab:my_label_2}, 
since these components are an artifact of the missing orthogonality between the stable 
and the unstable localized magnons. 

For the case $i_0=1$ it can be seen that the components are evenly distributed between 
the stable and unstable magnons (bigger orange and blue points in \figref{fig-kago-decomp} (right)). 
This is in good agreement with the values $P_0=P_{1/2}$ in Tab.~\ref{tab:my_label_2}. 
Overall, the contributions of the localized magnons drop sharply with the distance to the excitation.
As already mentioned, components in a non-orthogonal bases are not to be understood as 
contributions to a normalized state. Nevertheless, such a decomposition still
provides a good (approximate) understanding of the localization effects of the dynamics. 
            
To demonstrate the connection between this representation and the dynamics, 
the following auxiliary states are employed
            \begin{align}\label{eq-loc_mag_decomp}
                \ket{i_0, \varphi_\tau} = \sum_{\vnu}x_{\tau,\vnu}\ket{\varphi_\tau, \vnu} + \delta_{\tau,2} \gamma_\xi\ket{\xi}
                \ ,
            \end{align}
where the unit cell label $\vnu_0$ of the initial state $\ket{i_0,\vnu_0}$ is omitted. 
The states represent the contribution of the respective magnon type $\tau$ 
in the initial state in the representation according to the basis $\mathcal{B}$. 
The component of the dispersive bands is taken into account by the state
            \begin{align}\label{eq-dispersiv_decomp}
                \ket{i_0, \psi_\varepsilon} = \sum_{n=0}^{D-3l^2}\beta_n\ket{\psi_\varepsilon,n} .
            \end{align}
The four introduced states form a decomposition of an initial state $\ket{i_0,\vnu_0}$, i.e.
            \begin{align}
                \ket{i_0,\vnu_0} = \sum_{\tau=0}^2\ket{i_0, \varphi_\tau} + \ket{i_0, \psi_\varepsilon}\ .
            \end{align}

            \begin{figure}[!ht]
                \centering
                \includegraphics[width=0.99\columnwidth]{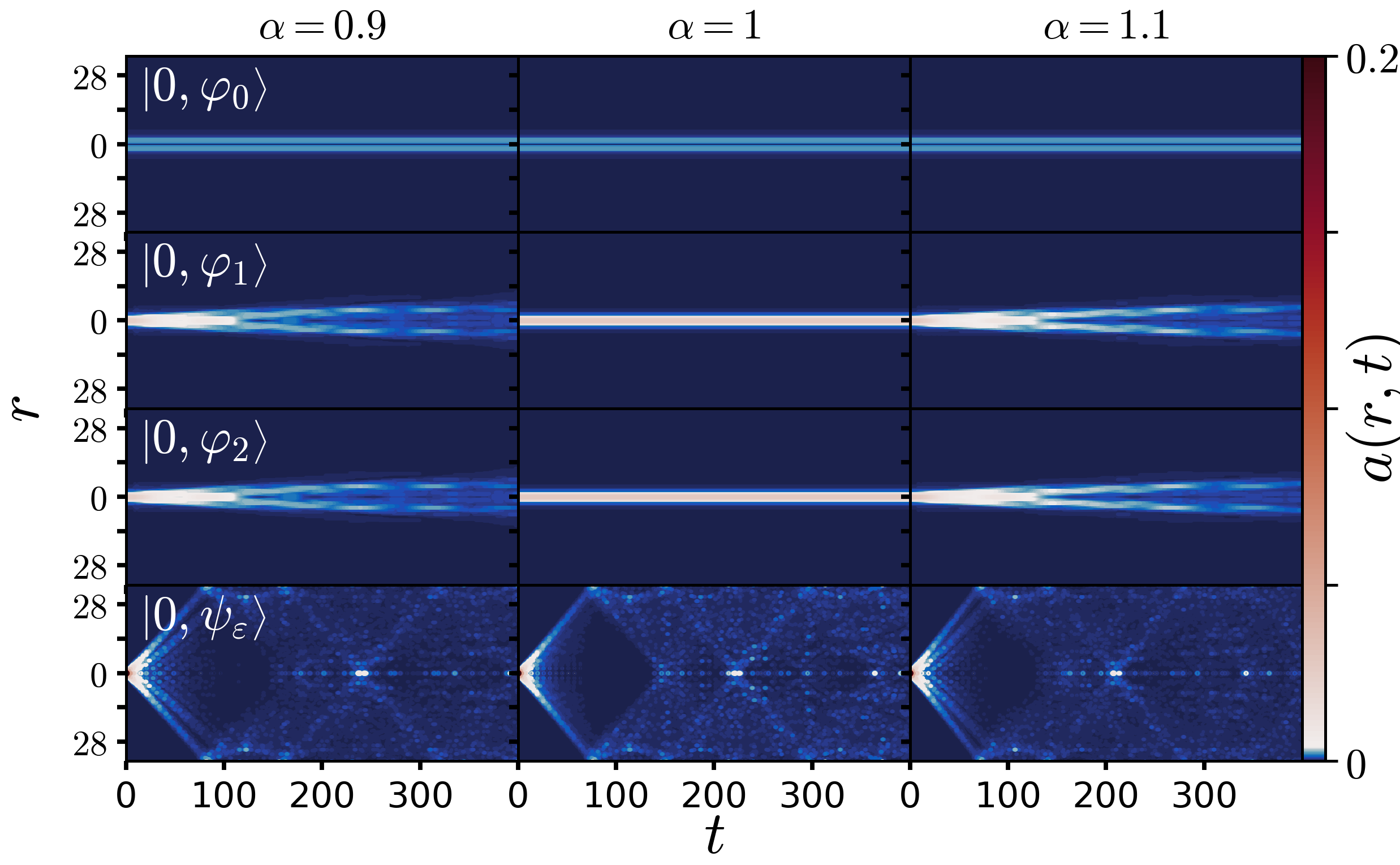}
                \caption{
                The local magnetization $a(r, t)$
                (\eqref{eq-absenk}) of the \KG with $l=36$ as
                a function of time and distance $r$ to the spin
                position $(i_0=0,\nu_0)$ for different values of
                $\alpha$ and different initial states, see
                Eqs.~\fmref{eq-loc_mag_decomp}
                and \fmref{eq-dispersiv_decomp}}.
                \label{fig-kago-approx}
            \end{figure}
            
Figure \ref{fig-kago-approx} shows the local magnetization $a(r,t)$ 
with these states as initial states for the case $i_0=0$ and different values of $\alpha$. 
To understand the figure, the missing orthogonality between the stable localized magnons 
and the unstable localized magnons has to be kept in mind. 
If this phenomenon is omitted, it can be confirmed that the different components of the observed 
dynamics (localized due to flat bands, dynamic due to almost flat bands, 
strongly dynamic due to dispersive bands) 
can be traced back to the dynamics of the respective states 
Eqs.~\fmref{eq-loc_mag_decomp} and \fmref{eq-dispersiv_decomp} for all values of $\alpha$. 
There is therefore a strong correlation between the localized magnons, 
their stability and the dynamics of the system. 
            
            \begin{figure}[!ht]
                \centering
                \includegraphics[width=0.99\columnwidth]{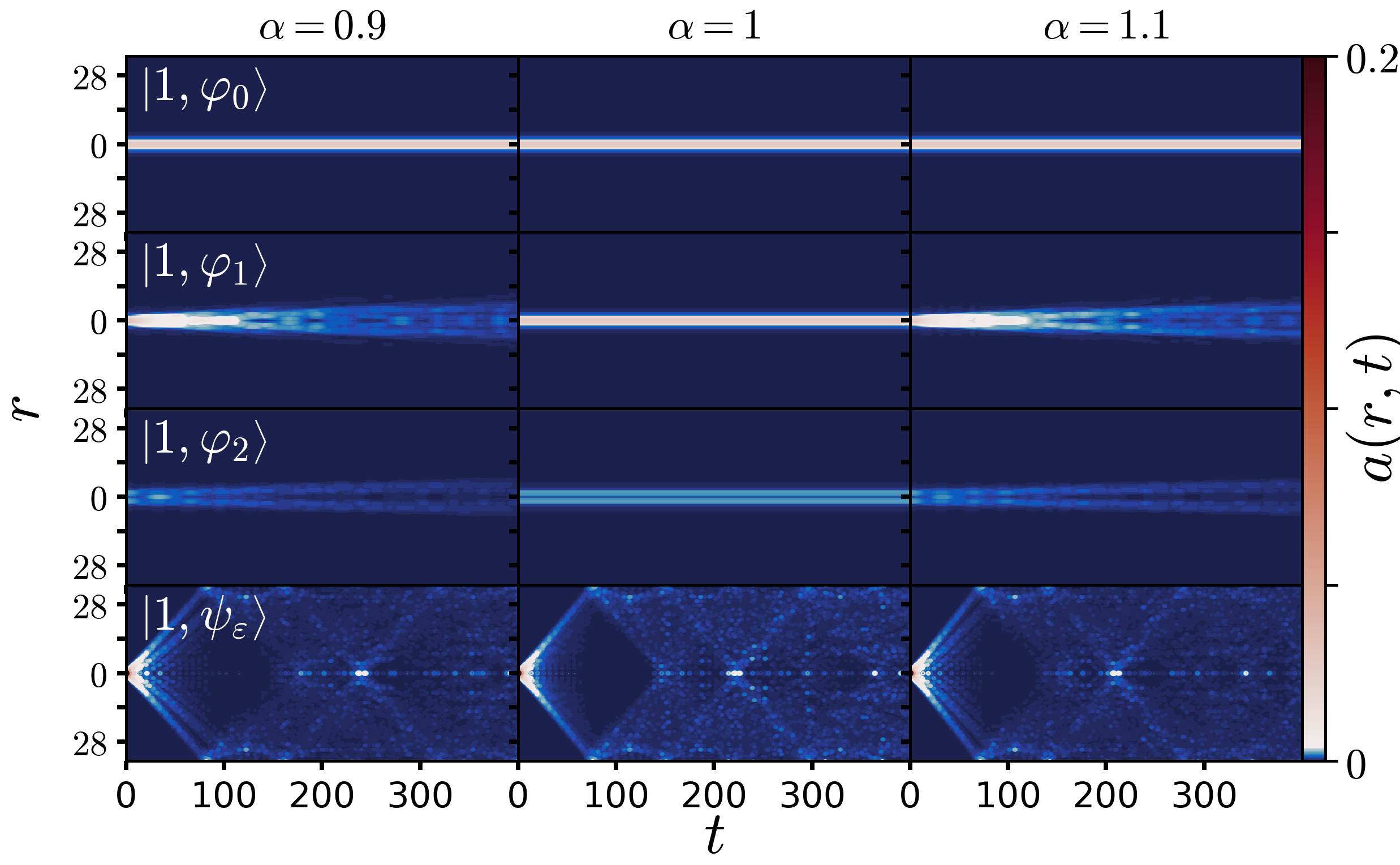}
                \caption{
                The local magnetization $a(r, t)$,
                \eqref{eq-absenk}, of the \KG with $l=36$ as
                a function of time and distance $r$ to the spin
                position $(i_0=1,\nu_0)$ for different values of
                $\alpha$ and different initial states, see
                Eqs.~\fmref{eq-loc_mag_decomp}
                and \fmref{eq-dispersiv_decomp}}.
                \label{fig-kago-approx-1}
            \end{figure}
            
The corresponding graphics of the dynamics for the case $i_0=1$ can be seen in \figref{fig-kago-approx-1}. 
Here, too, artifacts of the non-orthogonal representation can be observed. 
The contribution of unstable localized magnons with $\tau=2$ would be zero for this initial state 
if they were completed by an orthonormal basis to form a representation.  
The rest of the dynamics can again be viewed as a decomposition of the total dynamics, 
which shows the connection between localized magnons and the dynamics also for this initial state. 
            
With the aim of clarifying the phenomenon of the artifacts of
the non-orthogonal basis,
the initial states are next represented using the orthonormal
basis $\mathcal{B}^{\varphi_0}$, see \eqref{eq-ONB}.
This basis makes it possible to separate the stable from the
unstable localized magnon and to avoid the previous artifacts.
However, the distinguishability of the unstable localized
magnons among each other is lost.
The new partial states 
            \begin{align}\label{eq-decomp-1}
                \ket{i_0, \varphi_0^\prime} &= \sum_{\vnu}\ket{\varphi_0, \vnu}\braket{\varphi_0, \vnu}{i_0, \vnu_0} \ , 
            \end{align}
and
            \begin{align}\label{eq-decomp-2}
                \ket{i_0, \varphi_{1/2}^\prime} 
                &= \sum_{\tau=1}^2\sum_{\vnu}\ket{\varphi_\tau^\prime, \vnu}\braket{\varphi_\tau^\prime, \vnu}{i_0, \vnu_0} 
                \\
                &+ \delta_{\tau,2} \gamma_\xi\ket{\xi}
                \nonumber
            \end{align}
is defined. The partial state $\ket{i_0,\psi_\varepsilon}$ from the remaining base remains unchanged. 
            
\begin{figure}[!ht]
                \centering
                \includegraphics[width=0.99\columnwidth]{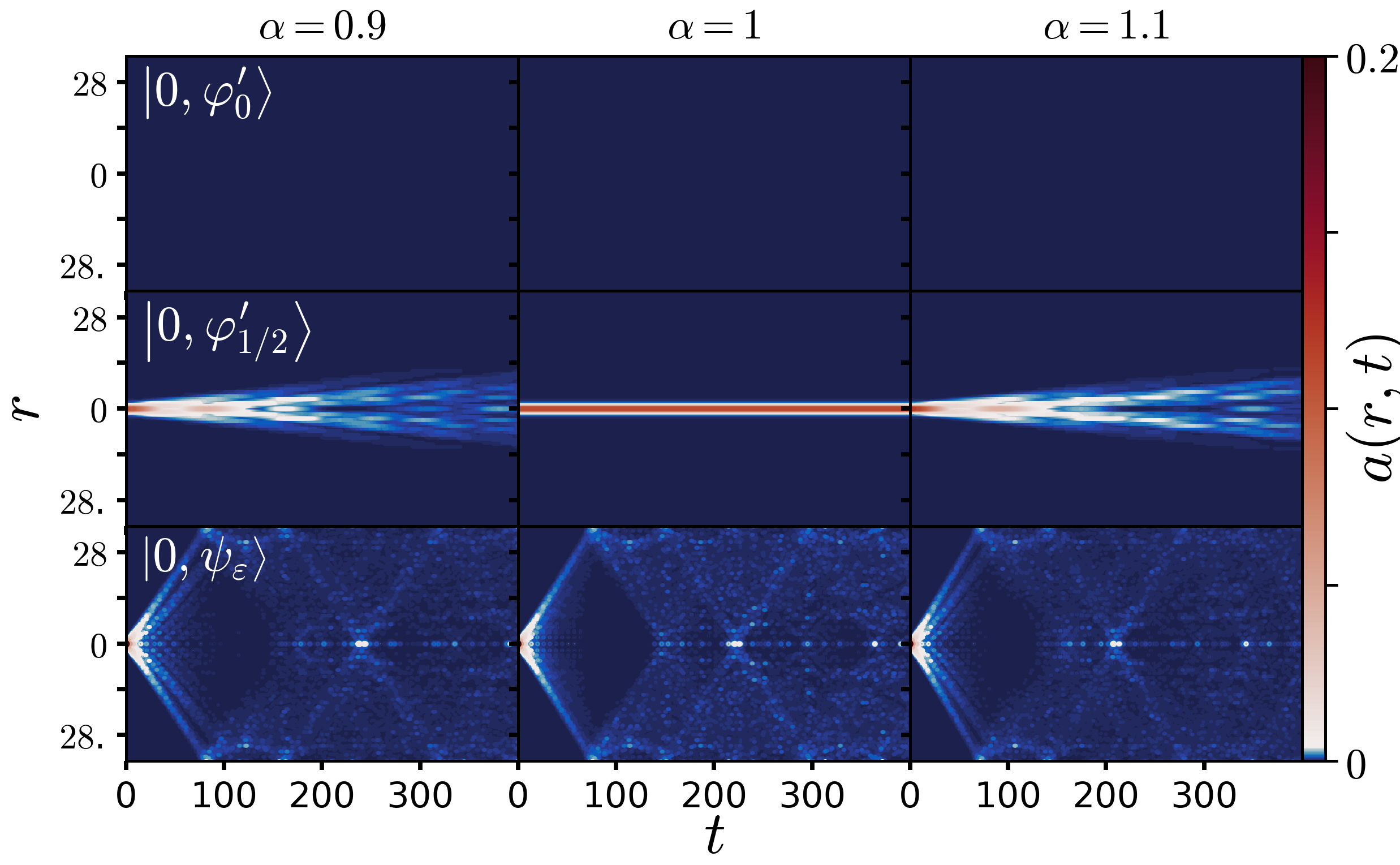}
                \caption{
                The local magnetization $a(r, t)$,
                \eqref{eq-absenk}, of the \KG with $l=36$ as
                a function of time and distance $r$ to the spin
                position $(i_0=1,\nu_0)$ for different values of
                $\alpha$ and different initial states, see
                Eqs.~\fmref{eq-dispersiv_decomp},
                \fmref{eq-decomp-1}, and \fmref{eq-decomp-2}}. 
                \label{fig-kago-approx-0-1}
            \end{figure}

\begin{figure}[!ht]
                \centering
                \includegraphics[width=0.99\columnwidth]{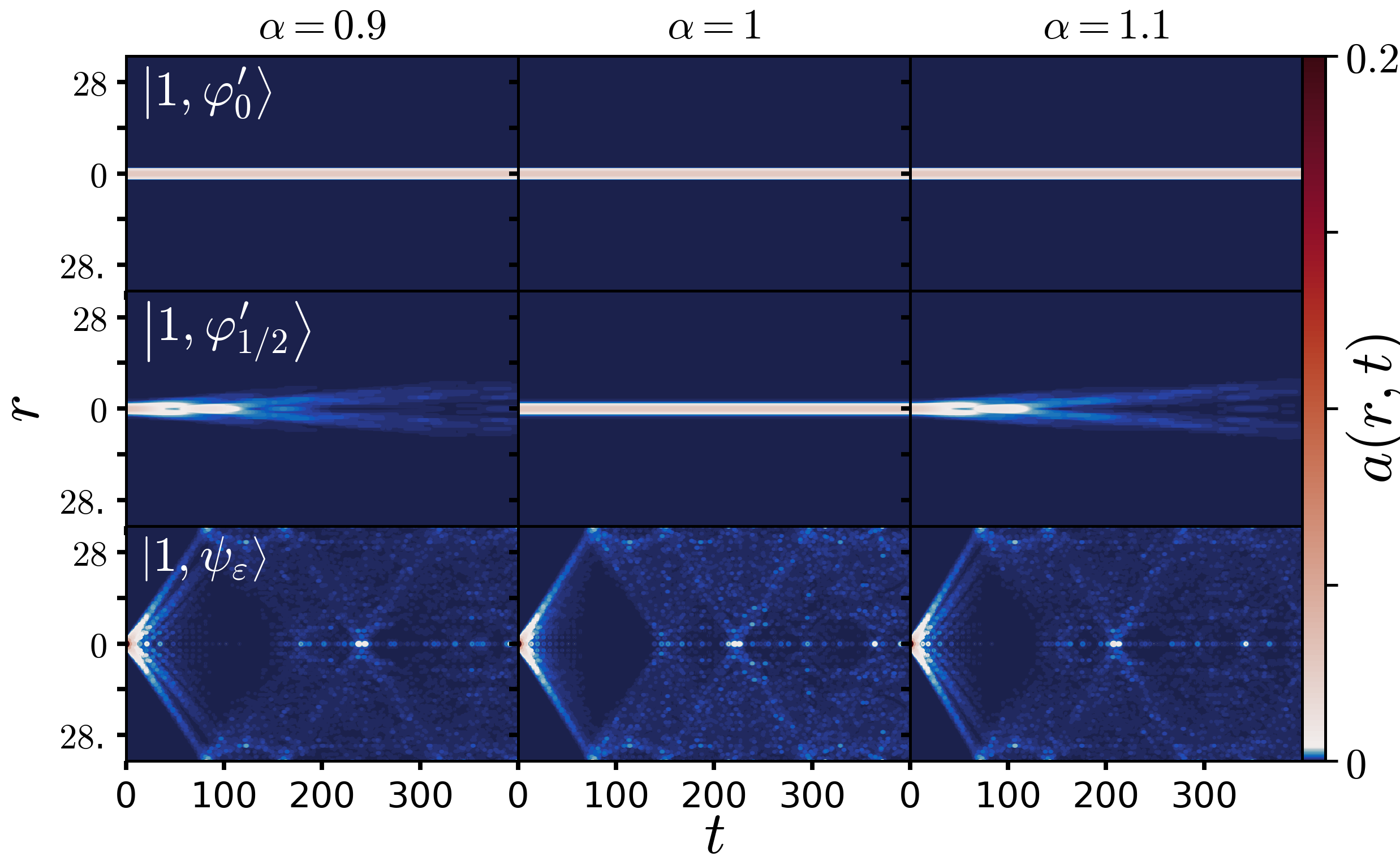}
                \caption{
The local magnetization $a(r, t)$, \eqref{eq-absenk}, of the \KG
with $l=36$ as a function of time and
distance $r$ to the spin position $(i_0=1,\nu_0)$ for different
values of $\alpha$ and different initial states,
see Eqs.~\fmref{eq-loc_mag_decomp} and \fmref{eq-dispersiv_decomp}.}
                \label{fig-kago-approx-1-1}
            \end{figure}

In the Figs.~\xref{fig-kago-approx-0-1} and
\xref{fig-kago-approx-1-1} the corresponding dynamics of
these states for both cases $i_0=0$ and $i_0=1$ is shown. 

For the initial state with $i_0=0$, \figref{fig-kago-approx-0-1}, it can be seen that (pseudo) 
parts of the stable localized magnons are no longer present and
the entire localized part of the
excitation comes from the unstable localized magnons, which decays for $\alpha\neq 1$. 
In the case $i_0=1$, see \figref{fig-kago-approx-1-1}, no artifacts can be detected either. 
However, this cannot be clearly verified by combining the unstable localized magnons. 
It can be summarized that the dynamics without artifacts can only be decomposed with 
the help of pairwise orthogonal partial states. As an outlook, the decomposition of the 
states using two further bases $\mathcal{B}^\varphi_1$ and $\mathcal{B}^\varphi_2$ 
which, analogous to the base $\mathcal{B}^\varphi_0$, 
explicitly contain the states of the unstable localized magnons, 
can already be suggested here.

\subsection{Long-term behavior of the \KG{s} as a function of system size}

This section examines the long-term behavior of the excitations as a function of system size. 
Specifically, it is discussed whether the localized behavior of the dynamics is an effect of the 
finite size of the system or whether this behavior is also maintained in the
limit $N\rightarrow\infty$.

\begin{figure}[!ht]
                \centering
                \includegraphics[width=0.99\columnwidth]{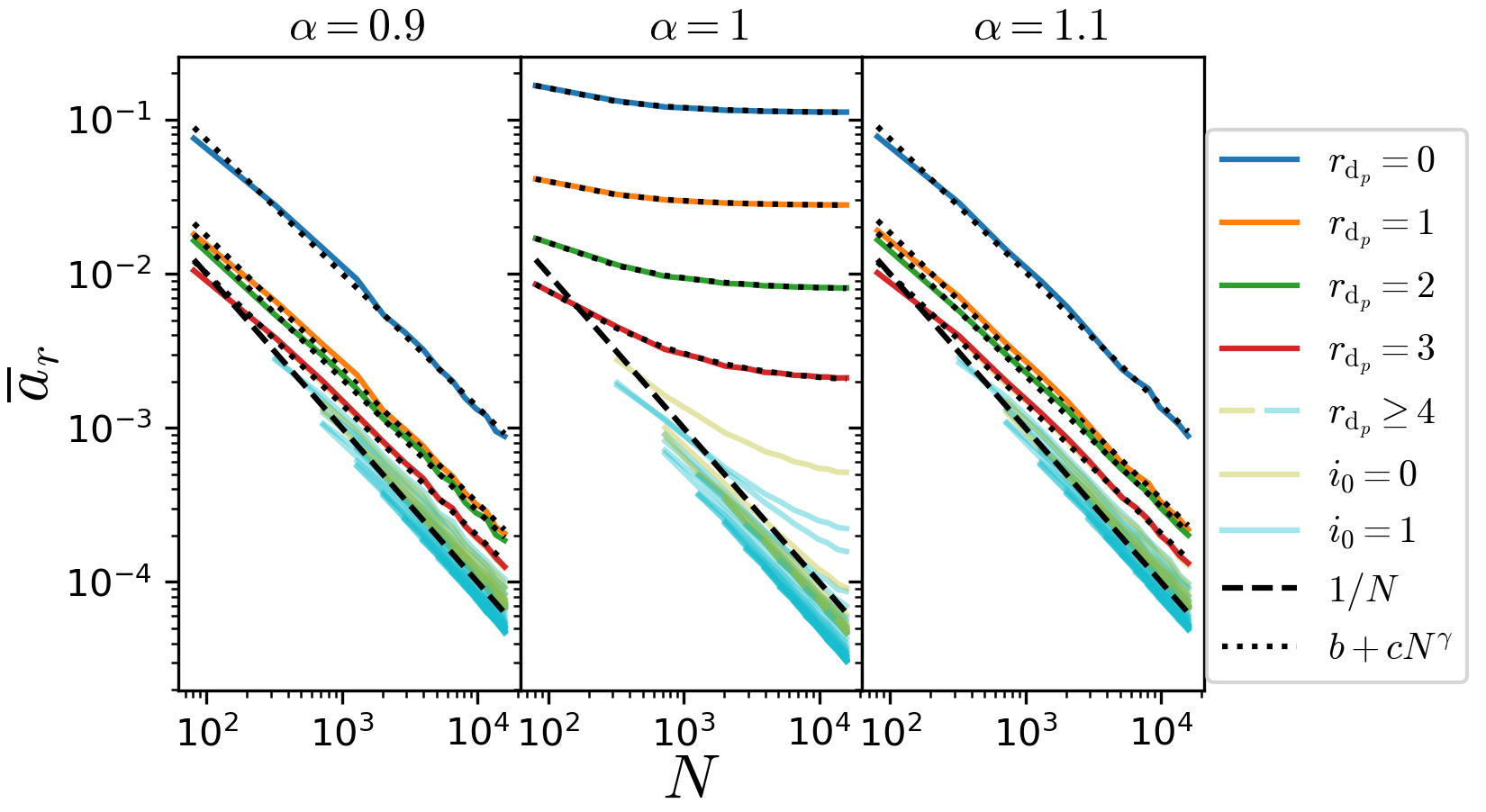}
                \caption{
                The time average of the magnetization $\overline{a}_r$
                of the \KG as a function of the system size $N$ related to an initial state 
                $\ket{0, \vnu_0}$ for different values of $\alpha$ over a late 
                time interval $[\Tilde{t}, \Tilde{t} + t_m]$
                with $\Tilde{t}=10^6$ and $t_m=2000$.}
                \label{lta-0-fig-kago}
            \end{figure}

\begin{figure}[!ht]
                \centering
                \includegraphics[width=0.99\columnwidth]{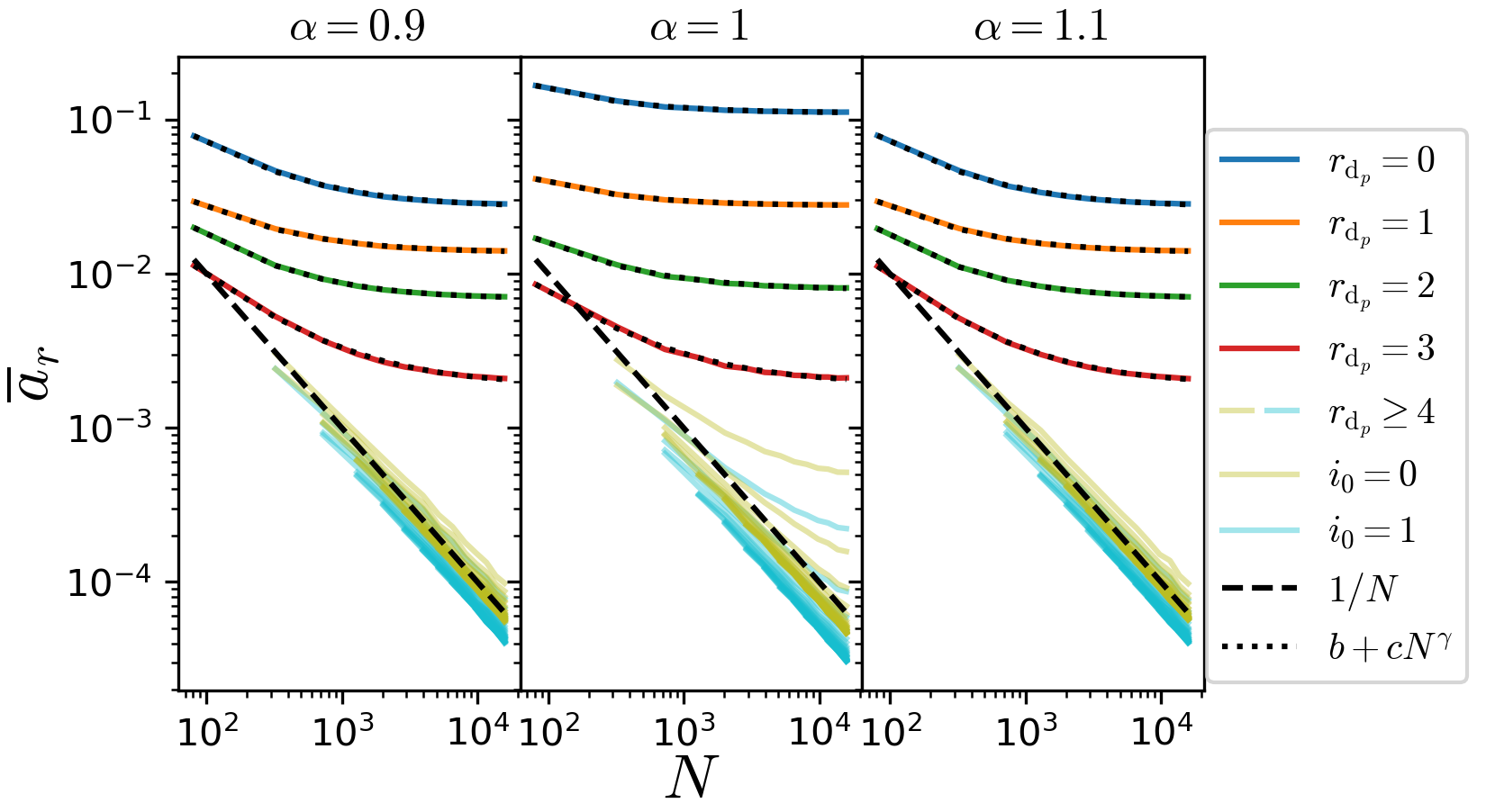}
                \caption{
                The time average of the magnetization $\overline{a}_r$
                of the \KG as a function of the system size $N$ related 
                to an initial state $\ket{1, \vnu_0}$ for different values of 
                $\alpha$ over a late time interval $[\Tilde{t}, \Tilde{t} + t_m]$
                with $\Tilde{t}=10^6$ and $t_m=2000$.}
                \label{lta-1-fig-kago}
            \end{figure}

For this purpose, Figs.~\ref{lta-0-fig-kago} and \ref{lta-1-fig-kago} 
show the time average of the local magnetization $\overline{a}_r$ 
at a late time $\Tilde{t}$ for spins 
with the topological distance $r=r_{\mathrm{d}_p}$ to the excitation. 
The integer values $r$ denote the $r$-th neighbors of the originally excited spin.  
The graphs show the results for different values of $\alpha$ as a function of the system size. 
It can be seen for the initial state with $i_0=0$, cf. \figref{lta-0-fig-kago},
that only in the homogeneous case some part of the excitation remains local. 
In this case, the magnetization of nearby spins does not seem to decay to zero 
with increasing system size. For values $\alpha\neq 1$, 
the excitations of all spins disappear with system size for this initial state. 

Following the argumentation in section \ref{sec-kago-dynamic} 
a model can be introduced that approximates $a^m_{r}(N)$ as
            \begin{align}
                a^m_{r}(N) = b_{r} + c_{r} N^{\gamma_{r}}
                \ .
            \end{align}
This model is based on the observation that for the case without remanent magnetization, 
i.e., $b_{r}=0$, 
the time-averaged excitation $\overline{a}_r$ shows a linear behavior depending on the system size 
in a doubly-logarithmic plot. This results in a relationship of the form 
$\overline{a}\propto N^\gamma$. If a remanent component exists, 
it is assumed that this behavior can be described by a value of $b_{r}\neq 0$. 
The difference to the model in section \ref{sec-kago-dynamic} 
is that the slopes $\gamma_r$ of the straight lines in the double-logarithmic plot 
deviate significantly from the value $\gamma_r=-1$.
      
In the Figs.~\xref{lta-0-fig-kago} and \xref{lta-1-fig-kago}, the curves fitted to the model 
are drawn as black-dotted lines. For orientation purposes, the function $1/N$ 
is also shown. 
It is evident that the model curves fit the results well. 
Their curvature in the doubly-logarithmic plot therefore 
appears to be due to a constant (remanent) contribution $b_r$.   
When considering the results for an initial state with $i_0=1$, 
cf. \figref{lta-1-fig-kago}, 
one notices that for all values of $\alpha$ there also seems to be a 
remanent part in the thermodynamic limit. Again, the model
curves fit the results very well.
          
\section{Summary and discussion}

Observations made for the sawtooth chain such as the absence of 
equilibration in one-magnon space can be taken over for the \KG.
However, not only is the technical analysis much more complex
in two dimensions, the dynamics is also much richer due to possible 
interference effects on the lattice. 
The interpretation of the dynamics in terms of stable and unstable 
localized magnons proved to be very insightful even if such a decomposition 
is only approximate.

However, it is apparent that states which contain stable localized magnons 
cannot fully equilibrate. This finding extends to other subspaces with
multi-magnon flat bands that can e.g.\ be derived from combinations of one-magnon 
flat bands \cite{EcS:PRR25}. Our conclusions thus hold for larger parts of the full
Hilbert space.

It is a special feature of our $J_1$-$J_2$ \KG that 
in the entire parameter range
there are localized parts in the dynamics which can be described by stable localized magnons. 
Finally, numerical evidence could be collected that the localization
behavior of the dynamics 
is not an artifact of a finite system size.

Our findings can be easily generalized to many other quantum spin systems 
hosting flat bands such as the the square-kagome lattice, the pyrochlore lattice,
certain bilayer systems \cite{YHB:PRB25} as well as several other frustrated systems.
They all provide examples of non-generic, non-ergodic dynamics \cite{MHS:PRB20,JES:PRB23}.
The discussed phenomenon is neither restricted to systems of spins $s=1/2$ 
nor to spin systems at all, but also appears e.g.\ for Hubbard models with flat bands 
\cite{Mie:JPA91A,Mie:JPA91,Mie:JPA92B,Tas:PRL92,MiT:CMP93,GiB:PRB96,Tas:PTP98,BeL:IJMPB13,DRM:IJMPB15,MMS:PRB17,LAF:APX18,TDD:ZN20,DaD:NJP24}.
Related issues of ``unconventional spin transport in strongly correlated kagome systems" are discussed in \cite{KPK:PRB24}.

\section*{Acknowledgements}

We would like to thank our collaborator and friend Johannes Richter,
who passed away in May 2025, for many insightful discussions.
HS likes to thank Katarina Kar{l}'ov{\'a} for inspiring discussions.
This work was supported by the Deutsche Forschungsgemeinschaft DFG
(355031190 (FOR~2692); 397300368 (SCHN~615/25-2)) as well as 
(449703145 (SCHN 615/28-1)).


%

\end{document}